\documentclass[10pt,letterpaper]{article}
\usepackage[top=0.85in,left=2.75in,footskip=0.75in]{geometry}

\usepackage{amsmath,amssymb}

\usepackage{changepage}

\usepackage{graphicx}
\usepackage{subcaption}

\usepackage[utf8x]{inputenc}

\usepackage{textcomp,marvosym}

\usepackage{cite}
\usepackage{float}
\usepackage{verbatim}

\usepackage{nameref,hyperref}

\usepackage{microtype}
\DisableLigatures[f]{encoding = *, family = * }

\usepackage[table]{xcolor}

\usepackage{array}
\usepackage[justification=centering]{caption}

\usepackage{ragged2e}

\newcolumntype{+}{!{\vrule width 2pt}}

\newlength\savedwidth

\raggedright
\usepackage[aboveskip=1pt,labelfont=bf,labelsep=period,justification=raggedright,singlelinecheck=off]{caption}

\makeatletter
\renewcommand{\@biblabel}[1]{\quad#1.}
\makeatother

\usepackage{lastpage,fancyhdr,graphicx}
\usepackage{epstopdf}
\fancyheadoffset[L]{2.25in}
\fancyfootoffset[L]{2.25in}
\begin{document}
\vspace*{0.2in}

\begin{flushleft}
{\Large
\textbf\newline{Israel–Hamas war through Telegram, Reddit and Twitter} % Please use "sentence case" for title and headings (capitalize only the first word in a title (or heading), the first word in a subtitle (or subheading) and any proper nouns).
}
\newline
\\
 Despoina Antonakaki\textsuperscript{1,2*},
 Sotiris  Ioannidis\textsuperscript{1, 2}
%Name6 Surname\textsuperscript{2\ddag},
%Name7 Surname\textsuperscript{1,2,3*},
%with the Lorem Ipsum Consortium\textsuperscript{\textpilcrow}
\\
\bigskip
\textbf{1} Technical University of Crete, University Campus, Akrotiri, Chania, Greece

\textbf{2} Institute of Computer Science, Foundation for Research and Technology, Vassilika Vouton, Heraklion, Crete
\bigskip

% Insert additional author notes using the symbols described below. Insert symbol callouts after author names as necessary.
%
% Remove or comment out the author notes below if they aren't used.
%
% Primary Equal Contribution Note
%\Yinyang These authors contributed equally to this work.

% Additional Equal Contribution Note
% Also use this double-dagger symbol for special authorship notes, such as senior authorship.
%\ddag These authors also contributed equally to this work.

%\textcurrency Current Address: Dept/Program/Center, Institution Name, City, State, Country % change symbol to "\textcurrency a" if more than one current address note
% \textcurrency b Insert second current address
% \textcurrency c Insert third current address


% Group/Consortium Author Note
%\textpilcrow Membership list can be found in the Acknowledgments section.

% Use the asterisk to denote corresponding authorship and provide email address in note below.
* Corresponding author \\ 
E-mail: dantonakaki@tuc.gr, despoina@ics.forth.gr

\end{flushleft}

 \section*{Abstract}
 
The Israeli-Palestinian conflict started on 7 October 2023, have resulted thus far to over 48,000 people killed including more than 17,000 children with a majority from Gaza, more than 30,000 people injured, over 10,000 missing, and over 1 million people displaced, fleeing conflict zones. The infrastructure damage includes the 87\%  of housing units, 80\% of public buildings and 60\% of cropland 17 out of 36 hospitals, 68\% of road networks and 87\% of school buildings damaged. 
This conflict has as well launched an online discussion across various social media platforms, including Twitter and Reddit. Telegram was no exception due to its encrypted communication and highly involved audience. The current study will cover an analysis of the related discussion in relation to different participants of the conflict and sentiment represented in those discussion. To this end, we prepared a dataset of 125K messages shared on channels in Telegram spanning from 23 October 2025 until today. Additionally, we apply the same analysis in two publicly available datasets from Twitter containing 2001 tweets and from Reddit containing 2M opinions. 
We apply a volume analysis across the three datasets, entity extraction and then proceed to BERT topic analysis in order to extract common themes or topics. Next, we apply sentiment analysis to analyze the emotional tone of the discussions. Our findings hint at polarized narratives as the hallmark of how political factions and outsiders mold public opinion. We also analyze the sentiment-topic prevalence relationship, detailing the trends that may show manipulation and attempts of propaganda by the involved parties. This will give a better understanding of the online discourse on the Israel-Palestine conflict and contribute to the knowledge on the dynamics of social media communication during geopolitical crises.
%To the best of our knowledge, this is the first study on Telegram regarding the Israeli -Palestinian conflict of 2023. 

%\linenumbers

\section{Introduction}
\label{sec:intro}
The Israeli-Palestinian conflict has persisted for over a century, with notable events ranging from the United Nations' 1947 Partition Plan, through the Yom Kippur War in 1973, to the recent outbreak of hostilities between Israel and Hamas in October 2023.
The recent ongoing Israeli-Palestinian conflict continued to dominate discussions across social media platforms, particularly Twitter and Telegram, where communities engage in sharing perspectives, news, and narratives. This study aims to analyze the conversations surrounding this conflict by applying advanced analytical techniques, including Latent Dirichlet Allocation (LDA), BERTopic analysis, and sentiment analysis, on a dataset obtained from Telegram, along with publicly available datasets from Twitter and Reddit. 

Our dataset comprises of 125.054 messages collected from various Telegram channels (see table \ref{tab:channels}) dedicated to Palestinian issues, Israeli perspectives, and global reactions, as well as two publicly available datasets from Twitter containing 2001 tweets and from Reddit containing 2M opinions.
By employing topic and sentiment analysis, we uncover the key themes and topics being discussed, providing insight into the sentiments expressed by users across different contexts. This analysis will help in understanding the diverse narratives surrounding the conflict and how they evolve in response to ongoing events.

In recent months, the discussions on Telegram have been characterized by a mixture of advocacy, solidarity, and expressions of frustration, reflecting the complex dynamics of the Israeli-Palestinian situation. The sentiment analysis component of our study reveals the prevailing emotions within these discussions, ranging from hope and support to anger and despair. By analyzing this dataset, we aim to contribute to the understanding of public sentiment and discourse on Telegram related to the Israeli-Palestinian conflict.

Furthermore, our findings may inform policymakers and social media platforms about the nature of conversations occurring in these digital spaces, helping to navigate the challenges posed by misinformation, propaganda, and the need for constructive dialogue.

\paragraph{Contributions}
\begin{itemize}
\item 
\textbf{Dataset Compilation and Analysis.}The study presents a unique dataset of over 125K Telegram messages, collected between 2023 and 2025, focusing on discussions about the Israeli-Hamas conflict. Also we include two publicly available datasets from Twitter containing 2001 tweets and from Reddit containing 2M opinions.  Thesw datasets are analyzed for volume trends, sentiment, and topics using advanced techniques, which adds significant value to the understanding of how such discussions unfold in encrypted social platforms during geopolitical crises.
\item \textbf{Sentiment-Topic Prevalence Analysis} This paper contributes a detailed sentiment-topic prevalence analysis using LDA and BERTopic. It helps reveal how emotional tones (positive, neutral, negative) align with specific topics, such as humanitarian concerns, political divisions, or propaganda efforts during the conflict. Such sentiment-based topic analysis is crucial for identifying manipulation or polarized narratives
%\item First Study of Israeli-Palestinian Conflict on Telegram: The paper marks the first known large-scale analysis of the Israeli-Hamas war on Telegram, emphasizing its relevance as a communication platform for both advocacy and misinformation. This contribution highlights the dual role Telegram plays in activism and manipulation during conflict periods
%\item 
%\textbf{Analysis of Propaganda and Polarization.} By identifying key patterns in propaganda efforts on both sides of the conflict, the paper sheds light on how Telegram channels are used for disinformation and influence campaigns. This provides a valuable framework for understanding how public opinion is shaped in digital spaces, particularly in times of heightened conflict.
\item 
\textbf{Application of Advanced NLP Models.} The use of cutting-edge natural language processing (NLP) models like BERTopic and sentiment analysis models allows for nuanced analysis of textual data. This methodological contribution improves the quality of analysis by ensuring that both topics and emotional tones are accurately captured across a large dataset.
\item 
\textbf{ Influence on Policy and Social Media Governance.}The findings regarding how narratives are shared and shaped on Telegram during the war could influence future policies on social media governance. The paper's results on the effectiveness of content moderation and the spread of information (and misinformation) could inform social media platforms' strategies during conflicts.
\end{itemize}
        .

These contributions provide a comprehensive analysis of how the Israeli-Hamas conflict is discussed on Telegram, combining data-driven approaches with important geopolitical and social insights.
\begin{comment}
    
Upon acceptance, we will make our dataset and analytical tools publicly available to foster further research and discussion on this crucial issue.
\end{comment}

\begin{comment}
\begin{table}[h!]
\centering
\begin{tabular}{|l|r|}
\hline
\textbf{Channel Name} & \textbf{Total Messages} \\
\hline
AlQassamBrigades9 & 2277 \\
Aqsatvsat         & 58   \\
bigolivr           & 42   \\
Eyeonpalestine2    & 3295 \\
FreePalestine2023  & 1    \\
gazaalanpa         & 6582 \\
gazaenglishupdates & 31988\\
GazaNow            & 70   \\
haqqintel          & 3896 \\
Palestine2024      & 10   \\
palestineonline    & 1329  \\
palestineresistance& 2381 \\
PalestineSolidarityBelgium & 147  \\
PalestineUpdates   & 464  \\
PalestinianResistance & 19 \\
pal\_Online9       & 47066\\
resistancechain    & 5428 \\
StopGazaGenocide   & 119  \\
The\_Jerusalem\_Post& 7706 \\
TIMESOFGAZA        & 1745 \\
total & 114.623 \\
\hline
\end{tabular}
\caption{Total number of messages in various Telegram channels on 2024-11-01}
\label{tab:channels}
\end{table}
 \end{comment}

 \begin{table}[h!]
\centering
\begin{tabular}{|l|r|}
\hline
\textbf{Channel Name} & \textbf{Total Messages} \\
\hline
AlQassamBrigades9 & 2326 \\
Aqsatvsat         & 58   \\
Eyeonpalestine2    & 7075 \\
FreePalestine2023  & 1    \\
GazaNow            & 70   \\
PalestineSolidarityBelgium & 147  \\
PalestineUpdates   & 465  \\
PalestinianResistance & 19 \\
StopGazaGenocide   & 119  \\
TIMESOFGAZA        & 1745 \\
The\_Jerusalem\_Post& 7907 \\
bigolivr           & 42   \\
gazaalanpa         & 6855 \\
gazaenglishupdates & 34688\\
haqqintel          & 4351 \\
pal\_Online9       & 49178\\
palestineonline    & 1778 \\
palestineresistance& 2455 \\
resistancechain    & 5775 \\
\hline
\textbf{Total}     & \textbf{125054} \\
\hline
\end{tabular}
\caption{Total number of messages in various Telegram channels on 2025-01-20}
\label{tab:channels}
\end{table}

\subsection{The landscape of Telegram, Twitter and Reddit on Palestine}
\label{landscapetelegram}
Telegram is one of the most important sources providing insights on the subject of the Israeli-Palestinian conflict and consequently a significant source of real-time data analysis; many participants and observers of the conflict use Telegram for quick communication but also considering its encrypted communication, given the restricted locations. Especially in groups or activities that are highly political or keen on violence, there is a preference for the use of Telegram for its encryption. It is this regulated space that impairs full participation and hence is useful for the analysis of unfiltered views among the population. 

However, the landscape of Telegram channels has significantly changed in the last few months, particularly due to the ongoing war in Palestine. Since the October 7 attack by Hamas, the use of Telegram on both sides has increased measurably among Israelis and Palestinians. It became one of the prominent channels through which information, propaganda, and live updates were passed around regarding the war. This in turn saw the emergence of new and much larger channels.

On the other hand, there have also been some remarkable limitations. Because Telegram's management was under increasing pressure with regard to content moderation, especially by Hamas, after some initial refusals to ban channels, it did implement measures to limit access to Hamas channels on platforms such as Google Play due to its app store guidelines. 

As of late October, there were reportedly channels affiliated with Hamas that were blocked or made less accessible, at least on cell-phones, but such efforts have tended to be quite half-hearted as many users still find ways to access this content - a very complex interplay between censorship and demand for such information.
However, the situation is fluid: some channels have appeared only recently, while others have been shuttered or buried deep due to these regulatory efforts. That would only point out that the current trend reflects the continuing challenge of content moderation on such platforms as Telegram during increased geopolitical tension.%  You may go through the articles from The Times of Israel and The Jerusalem Post for more details  \cite{Hamas_jpost, telegram_timesofisrael}.

Both Reddit and Twitter have been the most major sources of information dissemination and discussion on the Israeli-Palestinian conflict. Long used for real-time updates, debates, and information sharing, these platforms are open and accessible to a wide user base. However, both also tend to host highly polarized discussions, often leaning into ideological extremes, especially on sensitive and politically charged topics.

Because Reddit is an open discussion model, deep debates are going on, and subreddits range from Palestine to the Israeli-Palestinian conflict. These are very active, even since the escalation, with grassroots organizing going on and an international conversation happening there. The decentralized nature of Reddit sometimes creates difficulties in moderating extremist content and misinformation; some subreddits have been banned or highly moderated due to their ties to controversial or harmful content.

Twitter is similarly multifaceted: its real-time nature makes it a key space for breaking news, political discourse, and grassroots organizing. In the current war in Palestine, voices for and against take to Twitter to amplify their messages, updating live and organizing their support. While Twitter did try to rein in the proliferation of harmful content, including misinformation and hate speech, the avalanche of posts and hashtags made doing so particularly impossible in times of high conflict. Similar to Telegram, there are some increasing efforts to censor certain material or ban the accounts of extremist organizations, but many find ways through these restrictions while enforcement remains incomplete.

Ultimately, both the events on Reddit and Twitter reflect broader struggles concerning content moderation in an increasingly polarized political environment riddled with misinformation and shifting regulatory frameworks. Both remain fundamental spaces for expression, but they underpin the antithesis between free speech and content control in times of high geopolitical conflict.

\section{Related Work}

Similar works analyzing discourse on social media with regard to the Israel-Palestine conflict serve to highlight the dynamics of public sentiment and the forces of influence at play. 

In \cite{Baumgartner_Zannettou_Squire_Blackburn_2020} in a Telegram dataset of 27.8K channels and 317M messages from 2.2M unique users, the authors apply message frequency patterns and identifying key trends in channel activity over time. The dataset consists a wide-ranging corpus of messages from extremist groups, political movements, and discussions about cryptocurrency among other subjects. This makes it a rich source for the analysis of disinformation, scams, and political protests. Although this dataset was collected using snowball sampling, it is very comprehensive because data collection was a long-term process. The paper argues that the dataset will contribute significantly to research on computer-mediated communication and the use of platforms like Telegram in various global contexts.
The study in \cite{abdalla2023narratives} examines the process of narrative construction on social media during critical junctures, utilizing sentiment analysis as a way to show how public opinion is framed and subsequently manipulated by political groupings. In their work, they have established the importance of understanding the emotive subtext in the public discourse related to geopolitical conflicts.

In \cite{ben2023sentiment}, the authors focus on trends of sentiment in Telegram channels in several escalations of the Israel-Palestine conflict, using several topic modeling techniques to find topics discussed. Their results demonstrated that the level of conversation was polarized and some specific narratives were advanced in a loud tone by one or another party involved in this conflict.
 \begin{comment}
The work of %\cite{mohammed2023impact} 
examines social media's role in shaping public perceptions of the conflict with a special emphasis on misinformation and propaganda efforts. By utilizing more sophisticated machine learning methods, they trace patterns in shared content to show how misinformation campaigns link to trending topics.
 \end{comment}

 In \cite{amarasingam2021telegram}  they measure the impact of Europol's Action Days in 2018 and 2019 in disrupting jihadist networks on Telegram. It finds that while the first disruption, in 2018, had a minor effect, the second intervention significantly reduced jihadist activities on the platform, ushering in the long-term decrease in posts and the lifespan of the channels. Notably, as the study points out, following 2019, jihadists relocated increasingly to Twitter and Rocket.Chat. The findings indicate that it is the coordinated and sustained efforts, besides real-world policing, that bear fruit in containing online extremist activity.
 
In \cite{jongbloed2024analysing} they explore the quality of political discourse on social media, focusing on the Israel-Palestine conflict in Telegram channels. Drawing on Habermas's hypothesis of "disrupted public spheres," it examines how Telegram maintains low-quality, emotionally charged, and biased discourse. By applying a combination of machine learning and Critical Discourse Analysis, results show that Telegram messages of

%It is in this context that %\cite{jabar2023telegram}  investigate the potential for Telegram to be used as an information-spreading platform in the case of the Israeli-Palestinian conflict. Concretely, the authors focused on the peculiar features of Telegram that helped spread supportive and oppositional narratives and showed the duality of this platform for both activism and misinformation.
 Also in \cite{prucha2016and} they examine how groups like al-Qaida (AQ) and the self-designated Islamic State (IS) have used social media to advance their causes, particularly through da'wa, the religious obligation to recruit followers. While AQ initially led the use of online platforms, IS quickly overtook them, leveraging Twitter and later shifting to Telegram as a primary tool for recruitment, communication, and operational guidance. The article explores IS's sophisticated use of Telegram and its role in maintaining influence despite countermeasures on other social media platforms.
 
 Finally the paper in  \cite{guerra2024quantifying}  investigates the dynamics of extreme opinions on social media regarding the 2023 Israeli-Palestinian conflict, analyzing more than 450,000 Reddit posts from four subreddits. The research, using a lexicon-based unsupervised methodology, identifies peaks in extremism that are clearly linked to significant events in the conflict. The study explores the distribution of extremism scores across subreddits to derive insights about the spread of polarized sentiments and the challenges of measuring extremism in online discourse. It shows how social media analytics have the potential to deliver insights into real-world events and their outcomes in online discussion.

 Similar works on Twitter and Reddit Speaking explore the content of social media to
capture patterns or trends related to the ongoing conflict \cite{siapera2014tweeting,siapera2015gazaunderattack,deegan2018self} but apply only systematic statistical reviews, rather than data mining or sentiment analysis.

 In \cite@article{al2019multi} where they introduce a new analytic model that measures political public opinion about the Israeli-Palestinian conflict using Twitter data and includes both country-level and individual-level analyses of general attitudes, sentiment trends, opinion leaders, and ethnic groups.

In \cite{sharkar2024sentiment} they analyse a dataset from Kaggle with Reddit posts of 4,36,425 threads in total and apply sentiment analysis with 92.74\% accuracy with Sentiment Intensity Analyzer, while TextBlob, KNN, and SVM had 49.08\%, 53.08\%, and 86.13\%, respectively. Geopolitical stance analysis found 53\% of comments against Israel/Palestine, 27\% supporting Palestine, 17\% neutral or unclear, and 3\% supporting Israel.

\subsection{Background}
\label{Background}

\section{Dataset}
\label{dataset}

In order to achieve our analysis we obtain three datasets from Telegram(dataset A), Reddit (dataset B) and Twitter(dataset C). 
Regarding telegram, we acknowledge the fact that Telegram do not fully represent the userbase in Palestine during this period. Previous studies have shown that social media users belong to a specific age  cite{gayo2011limits}, social and ideology demographic group cite{preoctiuc2015studying}. This means that public opinion is not fully expressed through social media, especially in this crisis as described in section \ref{landscapetelegram}. The usage penetration on Telegram, at the start of 2024, Telegram is said to have close to 800 million monthly active users around the globe, easily placing it as one of the leading platforms in the world. The user base is especially strong in regions like Russia, India, and Brazil, which are countries with the highest amount of website traffic for the company. About 24\% of the total website visits come from Russia cite{SocialCapitalMarkets}.
In Palestine  Telegram boasts an impressive penetration rate at about 73\% of the region's internet users. By this level of usage, Telegram is among the top messaging platforms in Palestine. In that sense, from a local communications perspective, it provides a meaningful contact point. 

Giving the recent limitation on Twitter API, we could not query and retrieve the whole dataset regarding this period. Instead we use a publicly available dataset available here \cite{twitterGithubDataset} containing 2001 tweets, as described in table \ref{table:twitter_datasets}.

\begin{table}[h!]
\centering
\begin{tabular}{|c|c|}
\hline
\textbf{Dataset Name}              & \textbf{Number of Tweets} \\ \hline
gaza.csv                           & 501                      \\ \hline
israel.csv                         & 501                      \\ \hline
palestine.csv                      & 501                      \\ \hline
hamas.csv                          & 501                      \\ \hline
israel\_palestine\_conflict.csv     & 2001                     \\ \hline
 \end{tabular}
\caption{Twitter datasets related to the Israel-Palestine conflict from GitHub of user Rizqika Mulia Pratama}
\label{table:twitter_datasets}
\end{table}

Finally we also use a dataset from Reddit opinions, available here  \cite{redditkaggle}, with a total number of entries 2.134.070, containing comments from Reddit posts related to the current situation in Israel and Gaza.

\section{FAIR principles}
We guarantee that the presented dataset is compliant with the FAIR principles. This dataset is Findable because it has been publicly shared on the Zenodo platform; for this reason, it can be assigned a digital object identifier: {https://doi.org/10.5281/zenodo.14710657}. Secondly, the dataset is Accessible in that access to it is open to every person around the world and is in JSON format, which is a widely accepted data standard. Furthermore, the usage of JSON makes this dataset Interoperable because almost any programming language supports libraries for operating with JSON data. Finally, we share the whole dataset and its detailed description, along with links to the Telethon API documentation, allowing researchers to understand and operate the data. Please, for our work to remain Reusable, cite it when using this dataset.

\section{Methodology}

The data collection taking into consideration the limitations of reach to the userbase, as mentioned in section \ref{landscapetelegram}
span during a period of 9 years, with the oldest message on  2015-10-23 and the youngest message date 2024-10-24 , with a limited dataset of only 70313 messages covering all the channels mentioning Palestine.  
Having obtained our data set the next step is to apply a pre-processing  step on the of data like lowercasing, removing stop words, removing emojis etc. and then a volume analysis .

\subsection{Volume analysis}

\begin{figure}[H]
    \centering
    \begin{subfigure}{0.48\linewidth}
        \centering
        \includegraphics[width=\linewidth]{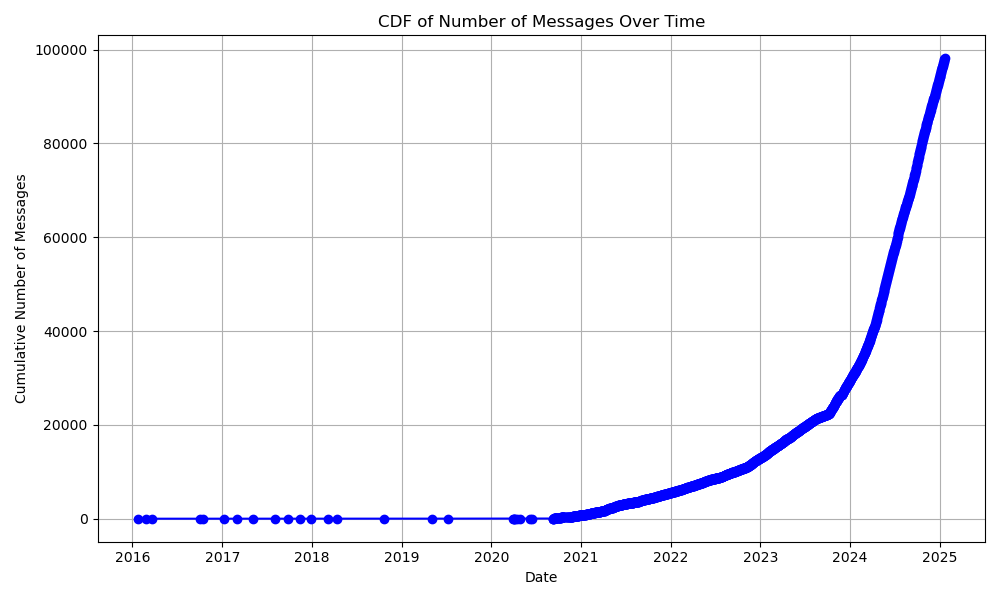}
        \caption{Overall volume of messages over time as CDF on Telegram.}
        \label{fig:all_volume_time}
    \end{subfigure}
    \hfill
    \begin{subfigure}{0.48\linewidth}
        \centering
        \includegraphics[width=\linewidth]{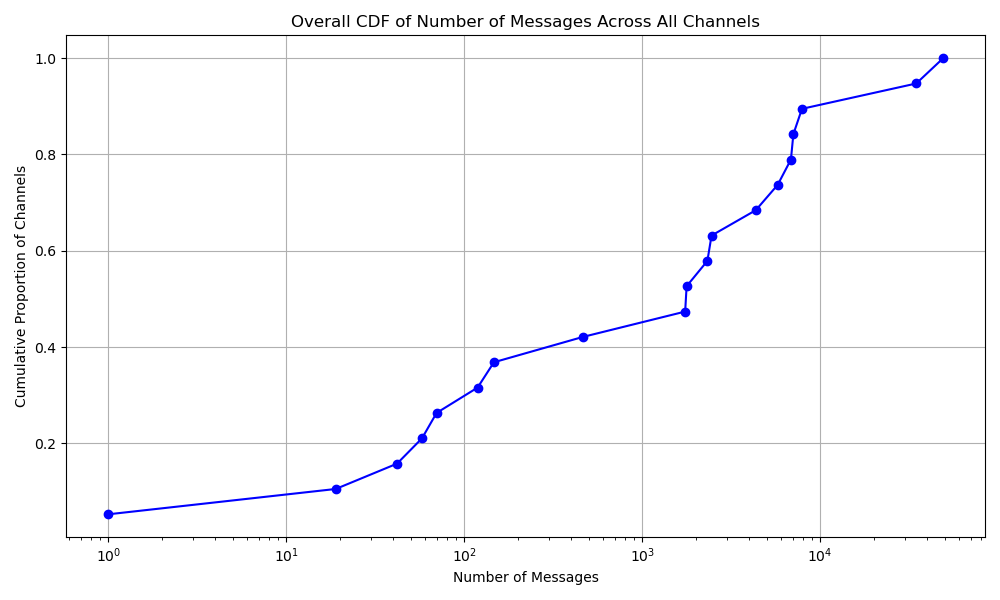}
        \caption{Overall volume across all channels as CDF on Telegram}
        \label{fig:all_volume_cdf}
    \end{subfigure}
    
    \vspace{0.5cm} % Adjust vertical space between rows

    \begin{subfigure}{0.48\linewidth}
        \centering
        \includegraphics[width=\linewidth]{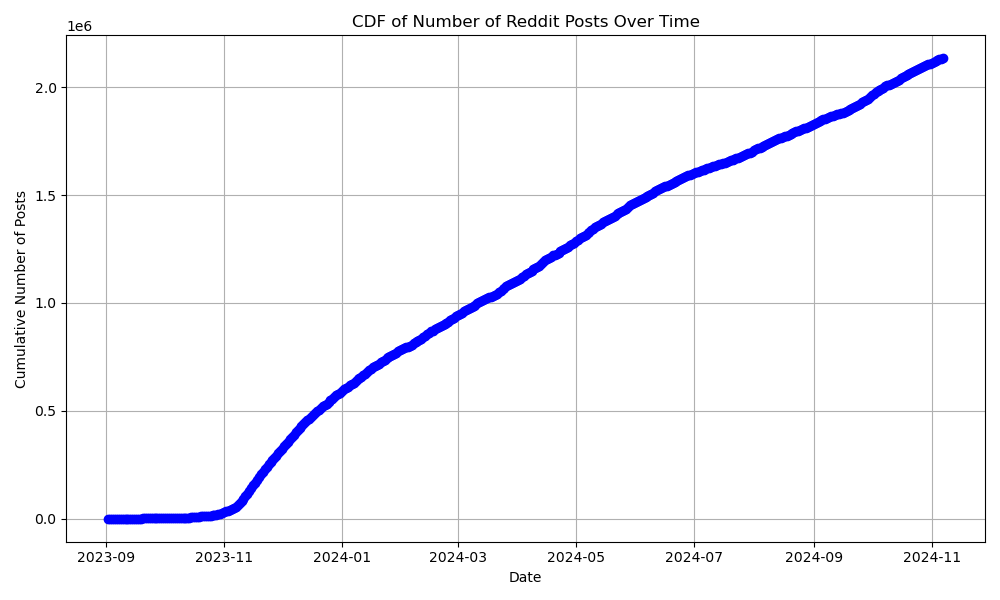}
        \caption{Overall volume in Reddit posts}
        \label{fig:reddit_volume}
    \end{subfigure}
    \hfill
    \begin{subfigure}{0.48\linewidth}
        \centering
        \includegraphics[width=\linewidth]{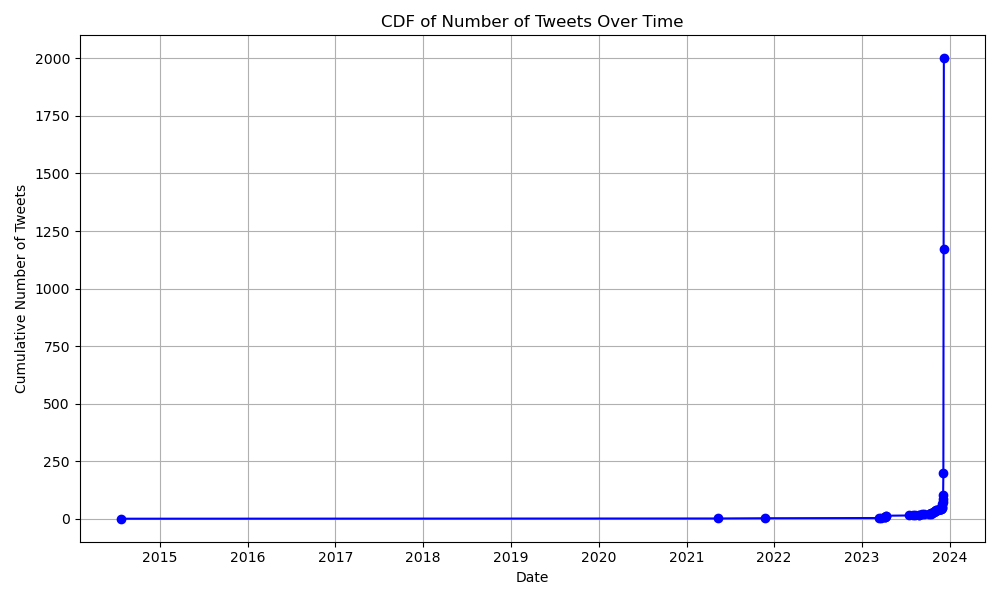}
        \caption{Overall volume in Twitter  posts}
        \label{fig:twitter_volume}
    \end{subfigure}
    
    \caption{Comparison of message volumes across platforms}
    \label{fig:volume_comparison}
\end{figure}

The first step before we proceed towards our analysis is the volume analysis of the messages in terms of time and channel distribution. As shown in  \ref{fig:all_volume_cdf} and \ref{fig:all_volume_time} the main volume of messages where initiated at the end of 2020 where the relationship between Hamas and Palestine  was characterized by ongoing tensions and developments in the broader context of Palestinian politics and regional dynamics including political divisions, normalization agreements, including as well the COVID-19 pandemic. In appendix in \ref{Add_volume} we show additional plot demonstrating the CDF of the volume of messages in the main channels .   

%/home/antonakd/TelegramCollector/topic/DataPreparation3.py 
\begin{comment}
    
 \begin{figure}[H]
    \centering
    \includegraphics[width=0.8\linewidth]{volume/cdf_messages_over_time.png}
    \caption{Overall volume of messages over time as cumulative distribution function}
    \label{fig:all_volume_time}
\end{figure}

\begin{figure}[H]
    \centering
    \includegraphics[width=0.8\linewidth]{volume/overall\_cdf\_messages.png}
    \caption{Overall volume in all channels plotted as cumulative distribution function}
    \label{fig:all_volume_cdf}
\end{figure}

\begin{figure}[H]
    \centering
    \includegraphics[width=0.8\linewidth]{volume/cdf_reddit_posts_over_time.png}
    \caption{Overall volume in reddit posts from dataset B}
    \label{fig:all_volume_cdf}
\end{figure}

\end{comment}

\begin{figure}[H]
    \centering
    \includegraphics[width=0.8\linewidth]{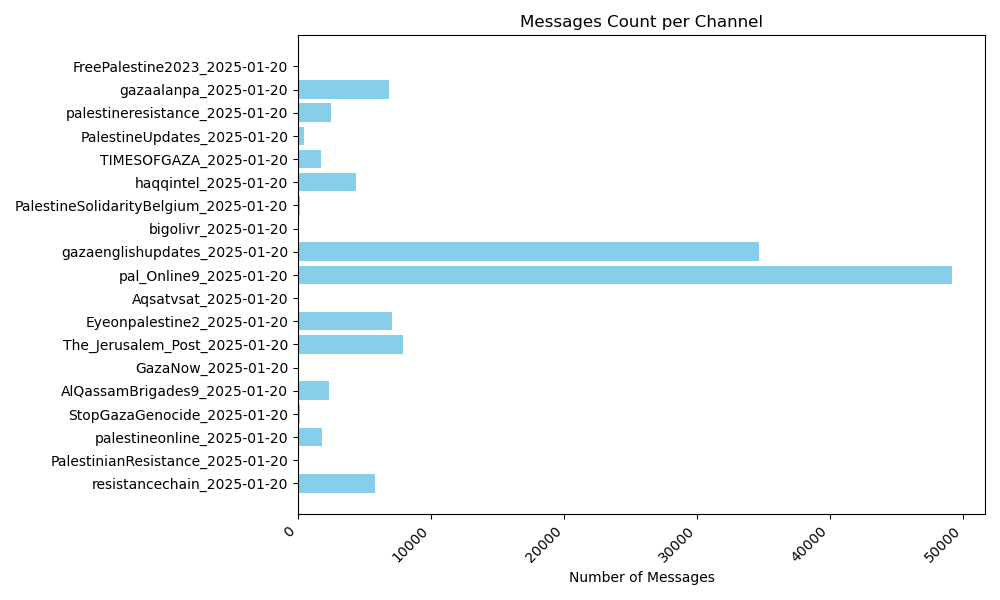}
    \caption{Barplot showing the distribution of messages per channel on Telegram}
    \label{fig:barplot}
\end{figure}

Additionally we show the barplot of the distribution of messages per channel in figure \ref{fig:barplot}

\begin{figure}[H]
    \centering
    \includegraphics[width=0.8\linewidth]{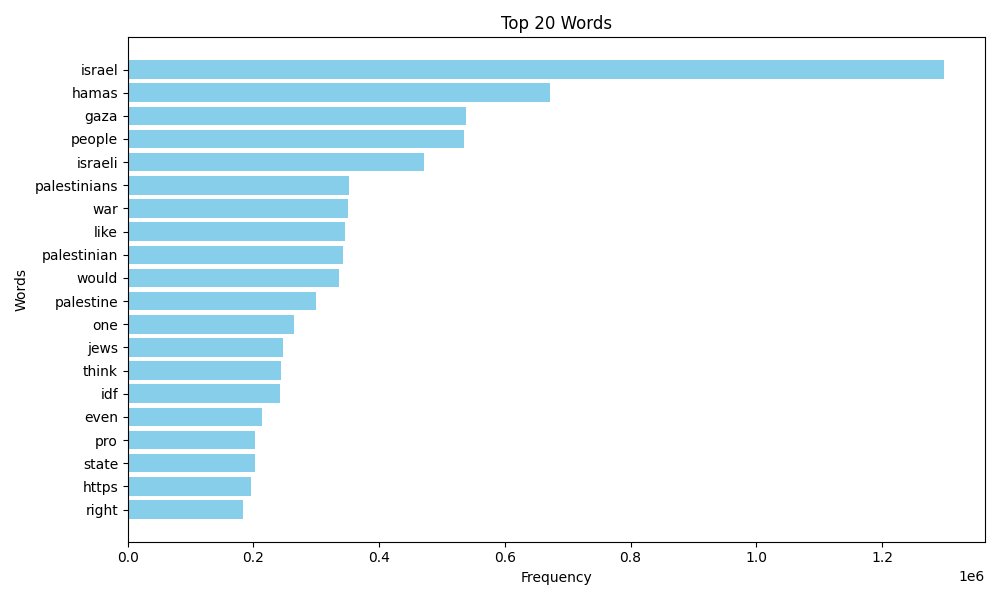}
    \caption{Barplot showing the distribution of messages per channel on Reddit}
    \label{fig:barplot_redditi_top}
\end{figure}

\begin{figure}[H]
    \centering
    \includegraphics[width=0.8\linewidth]{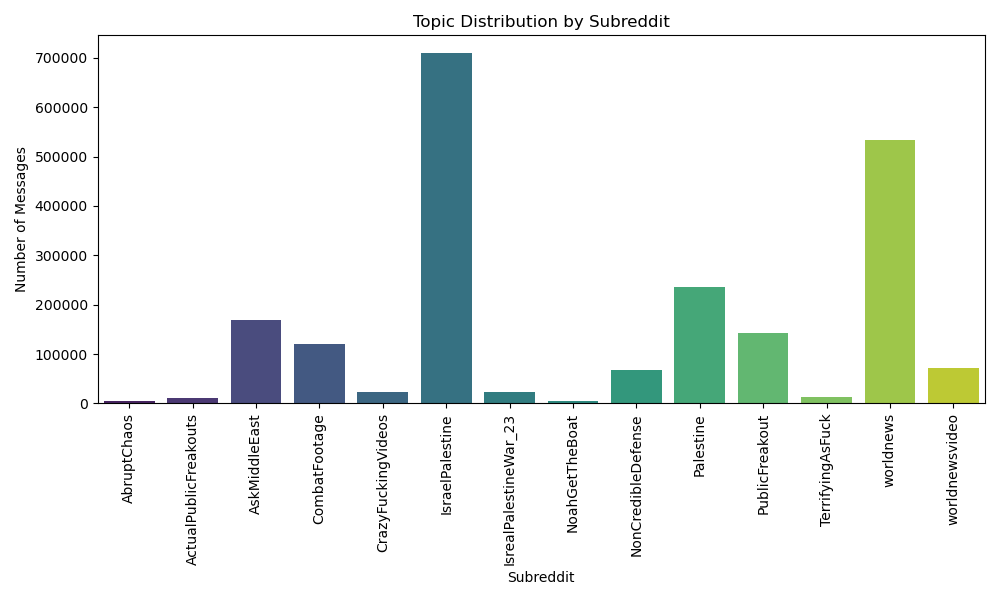}
    \caption{Messages per subreddit}
    \label{fig:barplot_reddit_distr}
\end{figure}

\begin{figure}[H]
    \centering
    \includegraphics[width=0.8\linewidth]{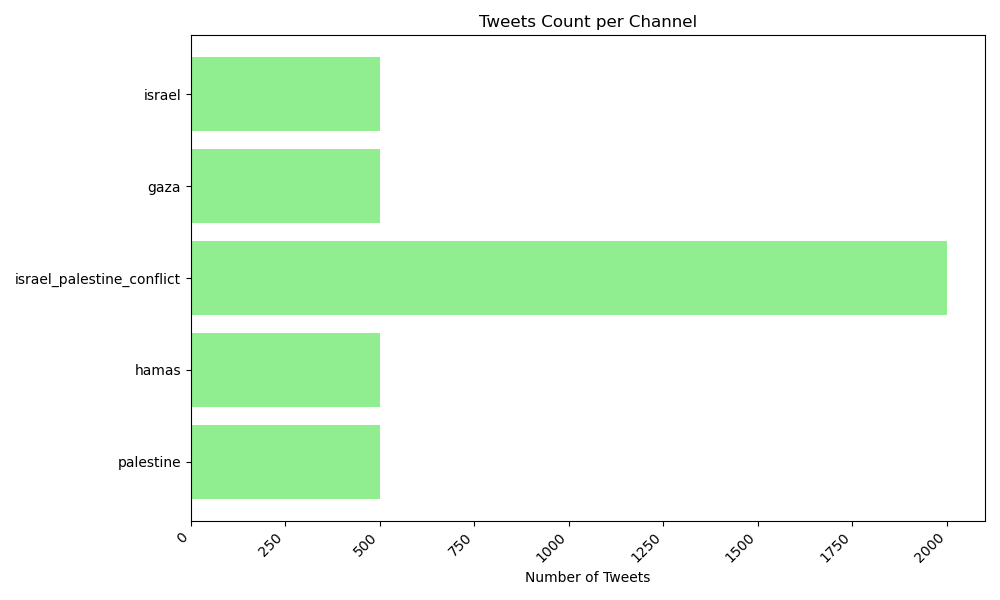}
    \caption{Barplot showing the distribution of tweets per dataset file}
    \label{fig:barplot_tweets}
\end{figure}

\begin{figure}[H]
    \centering
    % First Row of Subfigures
    \begin{subfigure}[b]{0.45\textwidth}
        \centering
        \includegraphics[width=\linewidth]{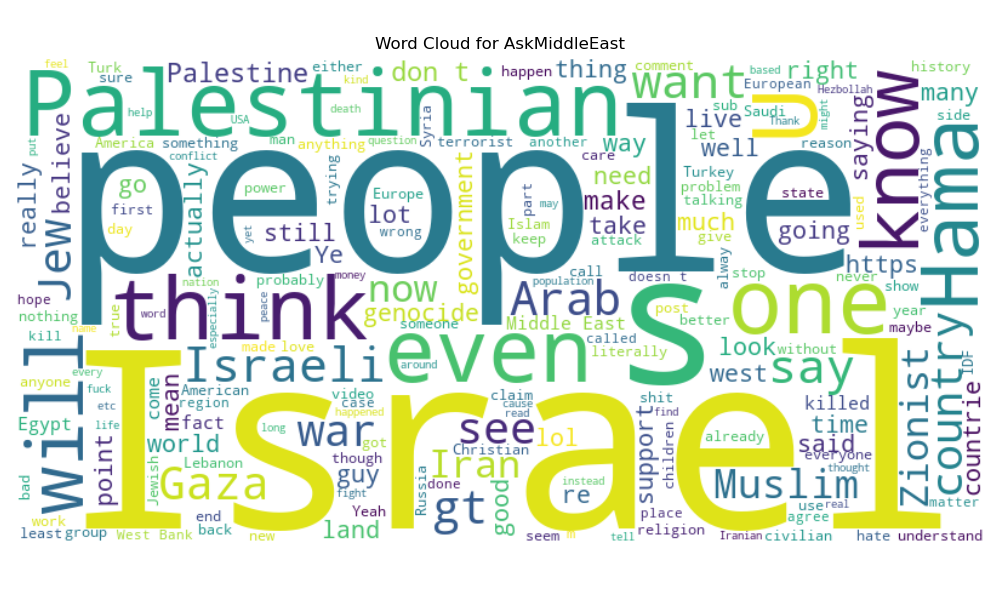}
        \caption{Word cloud for subreddit "Ask Middle East"}
        \label{fig:sub1}
    \end{subfigure}
    \hfill
    \begin{subfigure}[b]{0.45\textwidth}
        \centering
        \includegraphics[width=\linewidth]{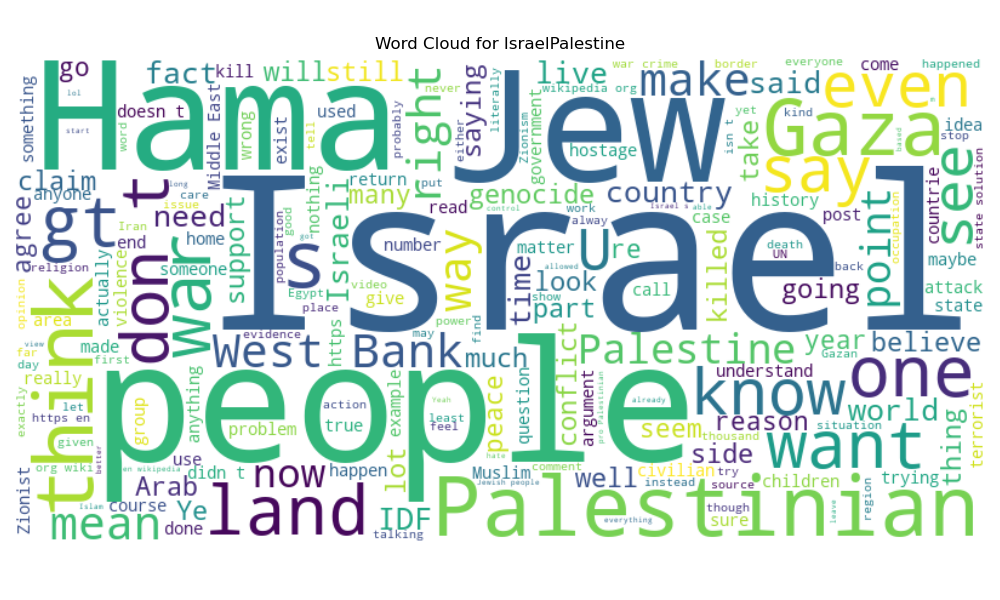}
        \caption{Word cloud for subreddit "Israel Palestine"}
        \label{fig:sub2}
    \end{subfigure}
    
    % Second Row of Subfigures
    \begin{subfigure}[b]{0.45\textwidth}
        \centering
        \includegraphics[width=\linewidth]{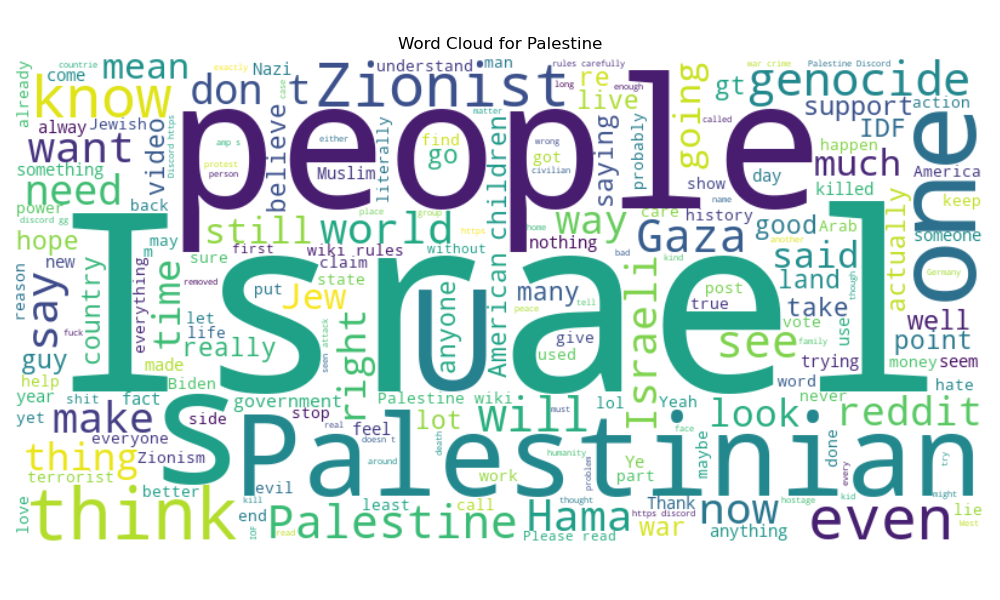}
        \caption{Word cloud for subreddit "Palestine"}
        \label{fig:sub3}
    \end{subfigure}
    \hfill
    \begin{subfigure}[b]{0.45\textwidth}
        \centering
        \includegraphics[width=\linewidth]{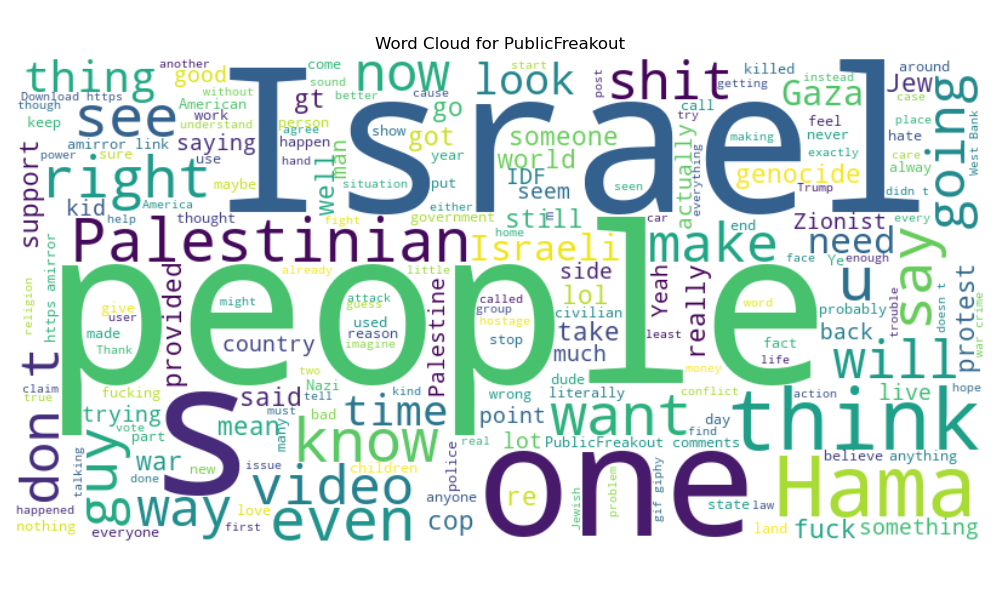}
                \caption{Word cloud for subreddit "Public Freak Out"}
        \label{fig:sub4}
    \end{subfigure}

    \caption{A 2x2 grid of subfigures showing various analyses.}
    \label{fig:4_subfigures}
\end{figure}

\subsection{Entity extraction} 
The first step of our analysis was to extract the main thematic entities the online discussion were centered on.
To achieve this, we process the messages from our dataset and extract the most common hashtags and words. Next we categorize these hashtags and words into predefined entities (such as people, events, institutions, and concepts) based on frequency and manually defined rule, resulting in the entities show in table with the entities and the number of mentions in all the dataset. 

\begin{table}[H]
\centering
\begin{tabular}{|l|r|}
\hline
\textbf{Entity} & \textbf{Frequency} \\ \hline
\#FreePalestine & 851 \\ \hline
\#Gaza & 563 \\ \hline
\#FreeThemAll & 259 \\ \hline
\#SaveSheikhJarrah & 214 \\ \hline
\#GazaUnderAttack & 187 \\ \hline
\#ShireenAbuAkleh & 152 \\ \hline
\#Palestine & 114 \\ \hline
\#FIFAWorldCup & 103 \\ \hline
\#ShutElbitDown & 83 \\ \hline
\#FIFAWorldCup2022 & 75 \\ \hline
\#SaveSilwan & 60 \\ \hline
\#GazaGenocide & 57 \\ \hline
\#SaveMasaferYatta & 55 \\ \hline
\#FreeAhmadManasra & 54 \\ \hline
\#WorldCup2022 & 49 \\ \hline
\#BoycottPuma & 46 \\ \hline
\#RaisePalestineFlag & 42 \\ \hline
\#NoToNormalization & 40 \\ \hline
\#Mawasi\_Massacre & 38 \\ \hline
israeli & 28238 \\ \hline
palestinian & 20592 \\ \hline
gaza & 20007 \\ \hline
occupation & 14214 \\ \hline
palestine & 9800 \\ \hline
forces & 7093 \\ \hline
israel & 6980 \\ \hline
palestinians & 6112 \\ \hline
people & 5093 \\ \hline
occupied & 5084 \\ \hline
strip & 4083 \\ \hline
resistance & 3910 \\ \hline
city & 3595 \\ \hline
children & 3585 \\ \hline
killed & 3567 \\ \hline
zionist & 3438 \\ \hline
solidarity & 3343 \\ \hline
https & 3214 \\ \hline
genocide & 3011 \\ \hline
camp & 3003 \\ \hline
\end{tabular}
\caption{Most common hashtags and words with their frequencies of appearance in the dataset on Telegram}
\label{tab:concepts}
\end{table}

\begin{figure}[H]
    \centering
    % First row of subfigures
    \begin{subfigure}{0.3\linewidth}
        \centering
        \includegraphics[width=\linewidth]{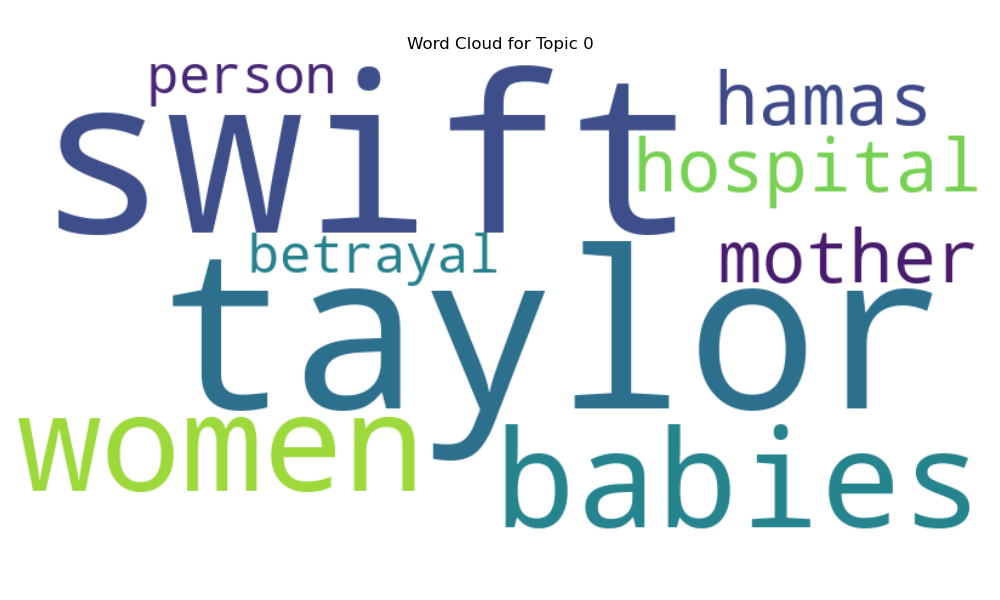}
        \caption{The entities contained in the 1st topic}
        \label{fig:topic1_secondtime}
    \end{subfigure}
    \hfill
    \begin{subfigure}{0.3\linewidth}
        \centering
        \includegraphics[width=\linewidth]{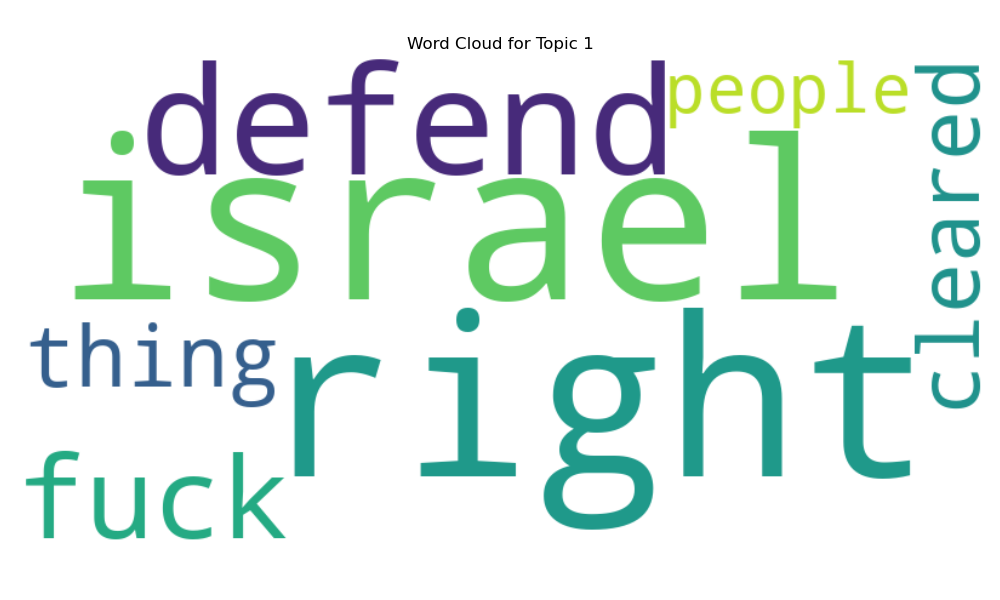}
        \caption{The entities contained in the 2nd topic}
        \label{fig:topic2}
    \end{subfigure}
    \hfill
    \begin{subfigure}{0.3\linewidth}
        \centering
        \includegraphics[width=\linewidth]{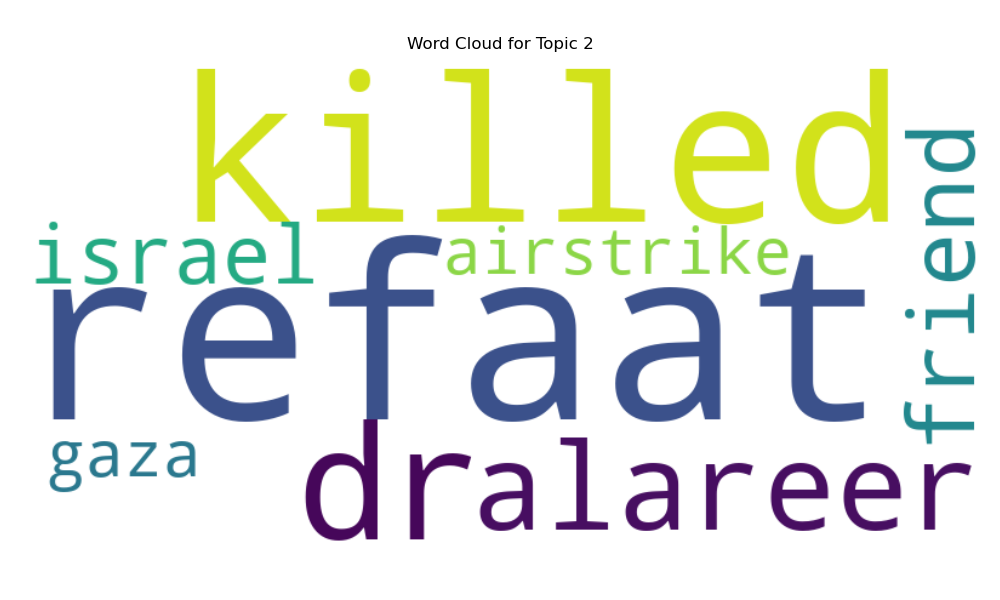}
        \caption{The entities contained in the 3rd topic}
        \label{fig:topic3_twitter}
    \end{subfigure}

    \vspace{0.5cm} % Add some vertical space between rows

    % Second row of subfigures
    \begin{subfigure}{0.3\linewidth}
        \centering
        \includegraphics[width=\linewidth]{twitter/word_cloud_topic_2.png}
        \caption{The entities contained in the 4th topic}
        \label{fig:topic4}
    \end{subfigure}
    \hfill
    \begin{subfigure}{0.3\linewidth}
        \centering
        \includegraphics[width=\linewidth]{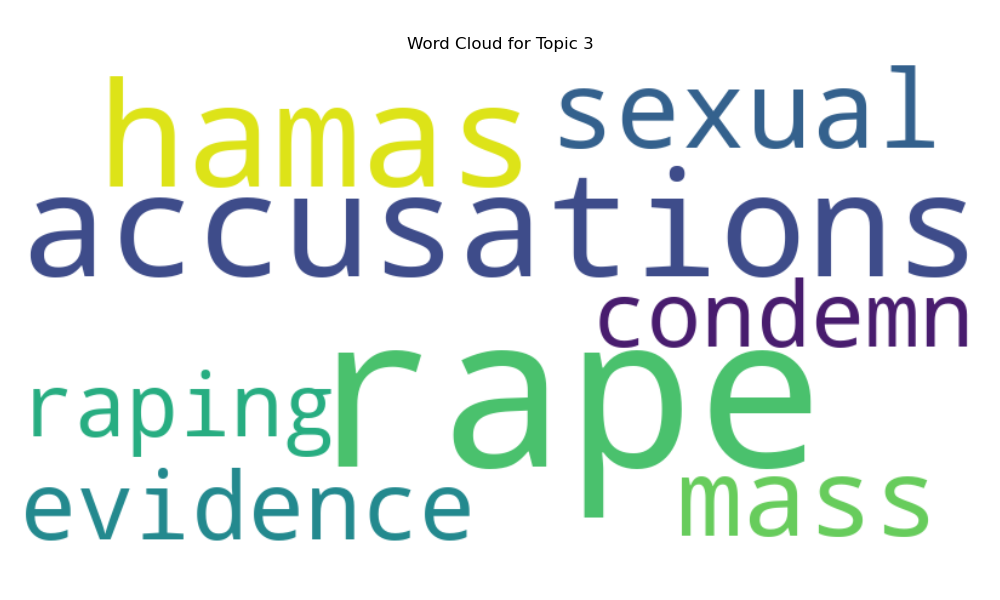}
        \caption{The entities contained in the 5th topic}
        \label{fig:topic5}
    \end{subfigure}
    \hfill
    \begin{subfigure}{0.3\linewidth}
        \centering
        \includegraphics[width=\linewidth]{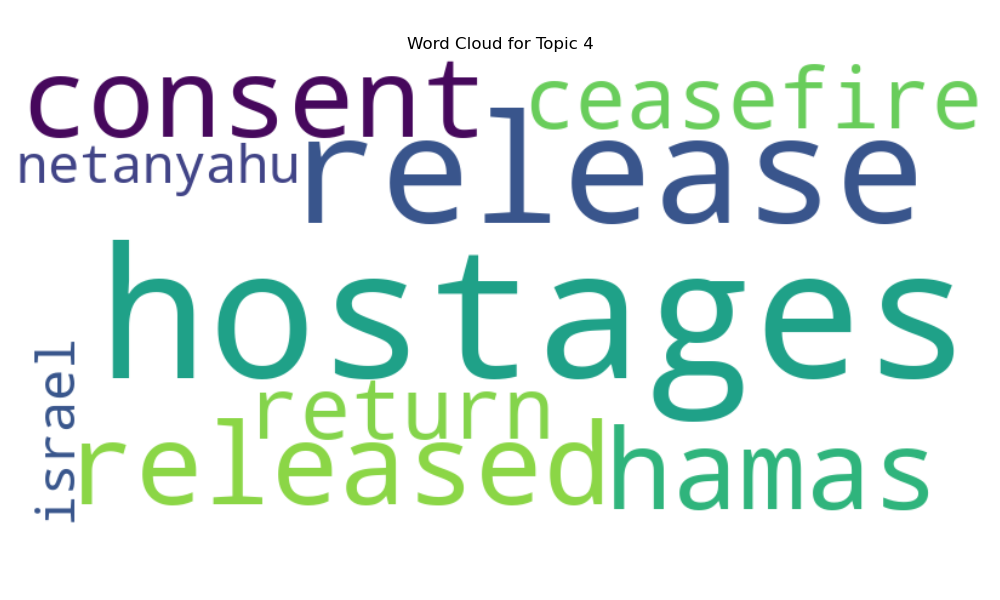}
        \caption{The entities contained in the 6th topic}
        \label{fig:topic6}
    \end{subfigure}

    \caption{Word clouds for each topic in the analysis on Twitter dataset}
    \label{fig:topics}
\end{figure}

\begin{figure}[H]
    \centering
    \begin{subfigure}{0.45\linewidth}
        \centering
        \includegraphics[width=\linewidth]{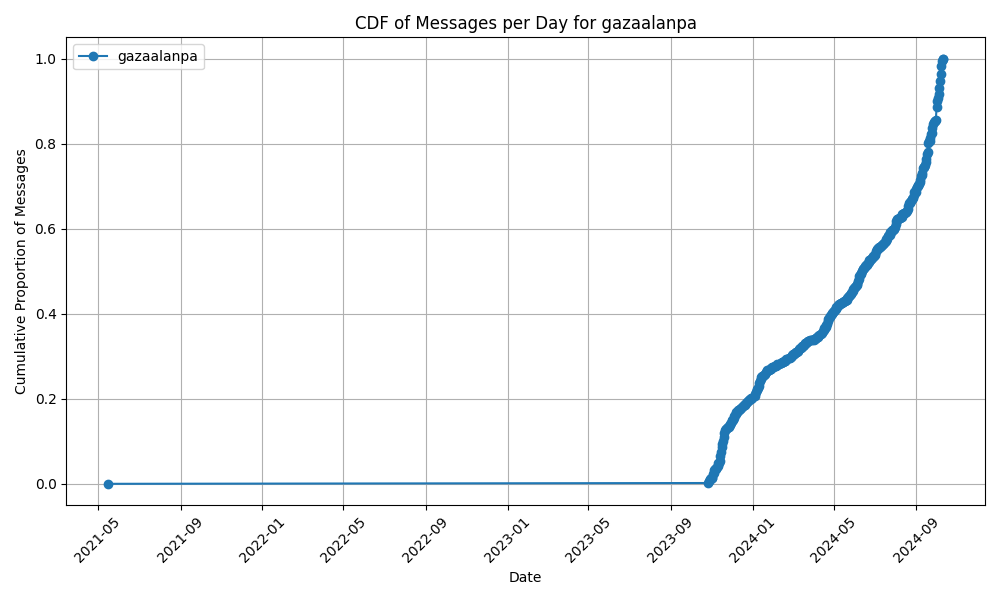}
        \caption{Cumulative distribution function of messages in 'GAZA NOW IN ENGLISH' channel}
        \label{fig:topic1}
    \end{subfigure}
    \hfill
    \begin{subfigure}{0.45\linewidth}
        \centering
        \includegraphics[width=\linewidth]{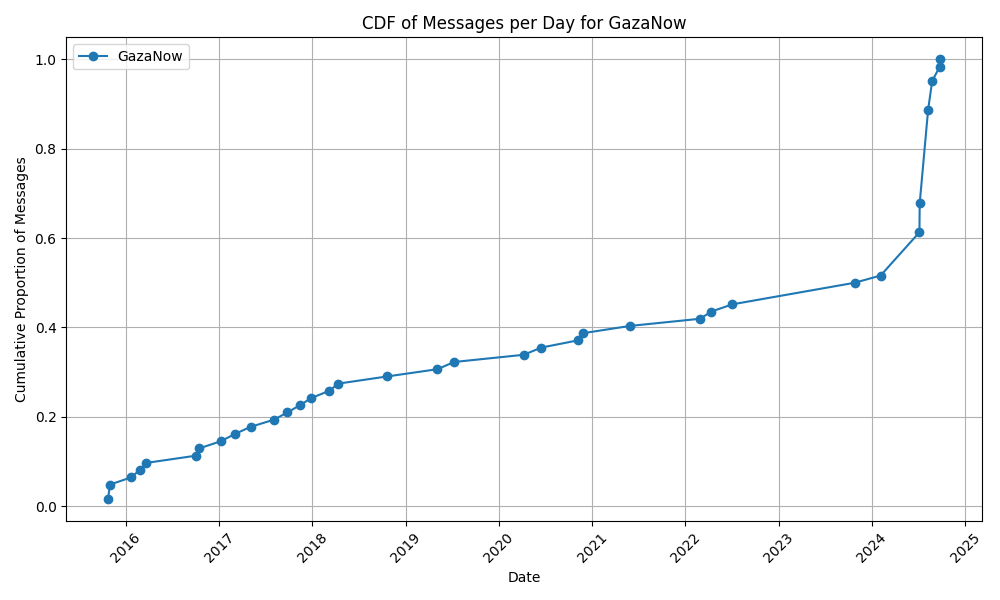}
        \caption{Cumulative distribution function of messages in 'GAZA IN ENGLISH' channel}
        \label{fig:GazaNow_cdf_plot}
    \end{subfigure}

    \vspace{0.5cm} % Add some vertical space between rows

    \begin{subfigure}{0.45\linewidth}
        \centering
        \includegraphics[width=\linewidth]{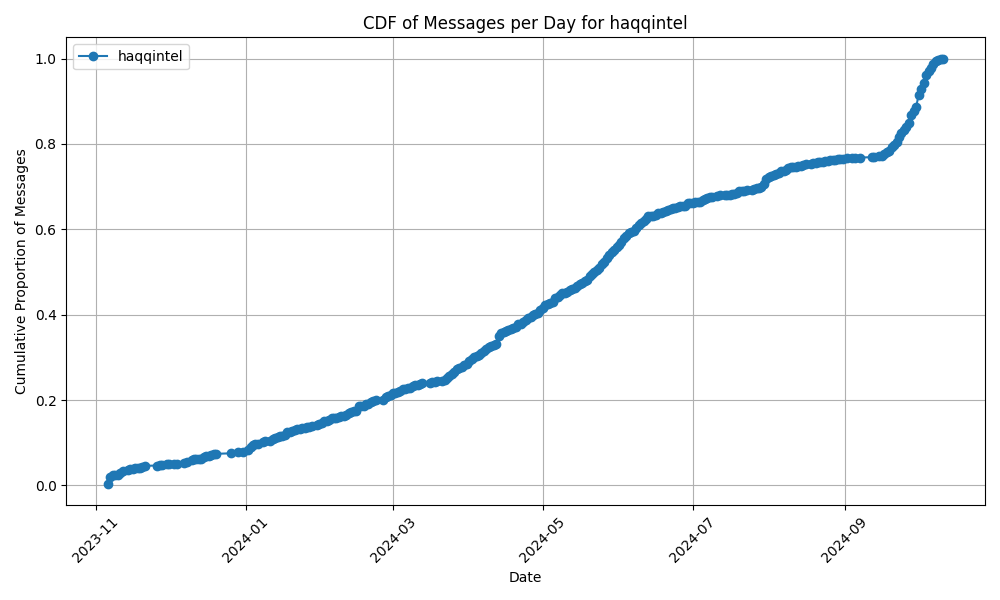}
        \caption{Cumulative distribution function of messages in 'Al Haqq' channel}
        \label{fig:haqqintel_cdf_plot}
    \end{subfigure}
    \hfill
    \begin{subfigure}{0.45\linewidth}
        \centering
        \includegraphics[width=\linewidth]{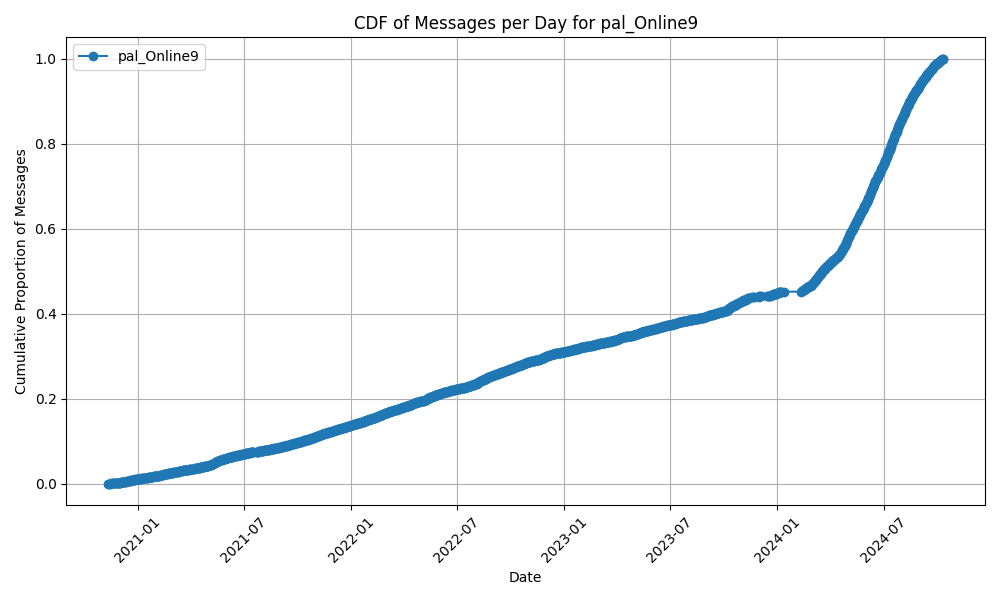}
        \caption{Cumulative distribution function of messages in 'Al Palestine Online' channel}
        \label{fig:pal_Online9_cdf_plot}
    \end{subfigure}

    \caption{Overall Caption for All Topics}
    \label{fig:all_topics}
\end{figure}

In figures \ref{fig:topic1}-\ref{fig:topic6}
we plot the most imporant enttiets prevaile in the Twitter dataset. 

\subsection{Topic Analysis}
\subsubsection{LDA}Towards the analysis of the topics discusssed in the corpus we apply state of the art methods for topic analysis including Latent Dirichlet allocation (LDA) and BertTopic analysis. LDA (Latent Dirichlet Allocation) and BERTopic were chosen as analytical techniques for topic modeling and sentiment analysis in this study because of their complementary strengths in capturing nuanced topics.

\begin{figure}[H]
    \centering
    \begin{subfigure}{0.45\linewidth}
        \centering
        \includegraphics[width=\linewidth]{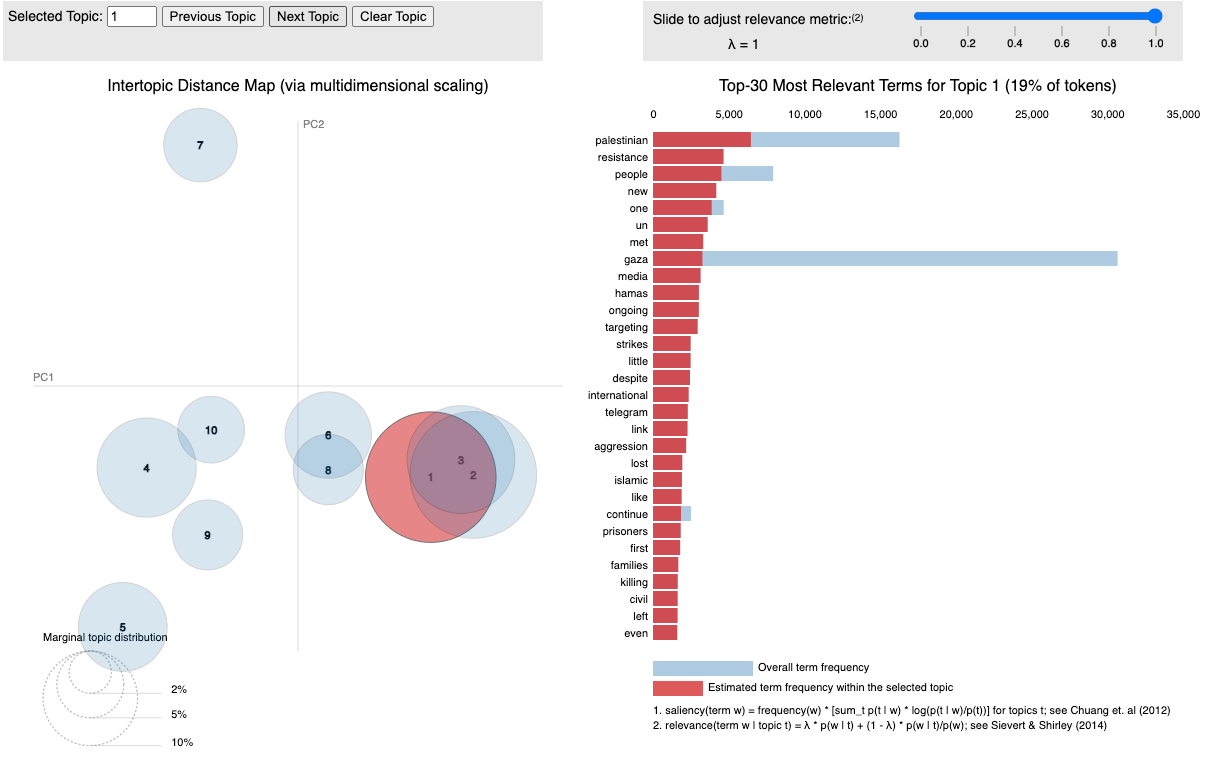}
	    \caption{Entities in topic 1}
        \label{fig:topic1_LDA}
    \end{subfigure}
    \hfill
    \begin{subfigure}{0.45\linewidth}
        \centering
        \includegraphics[width=\linewidth]{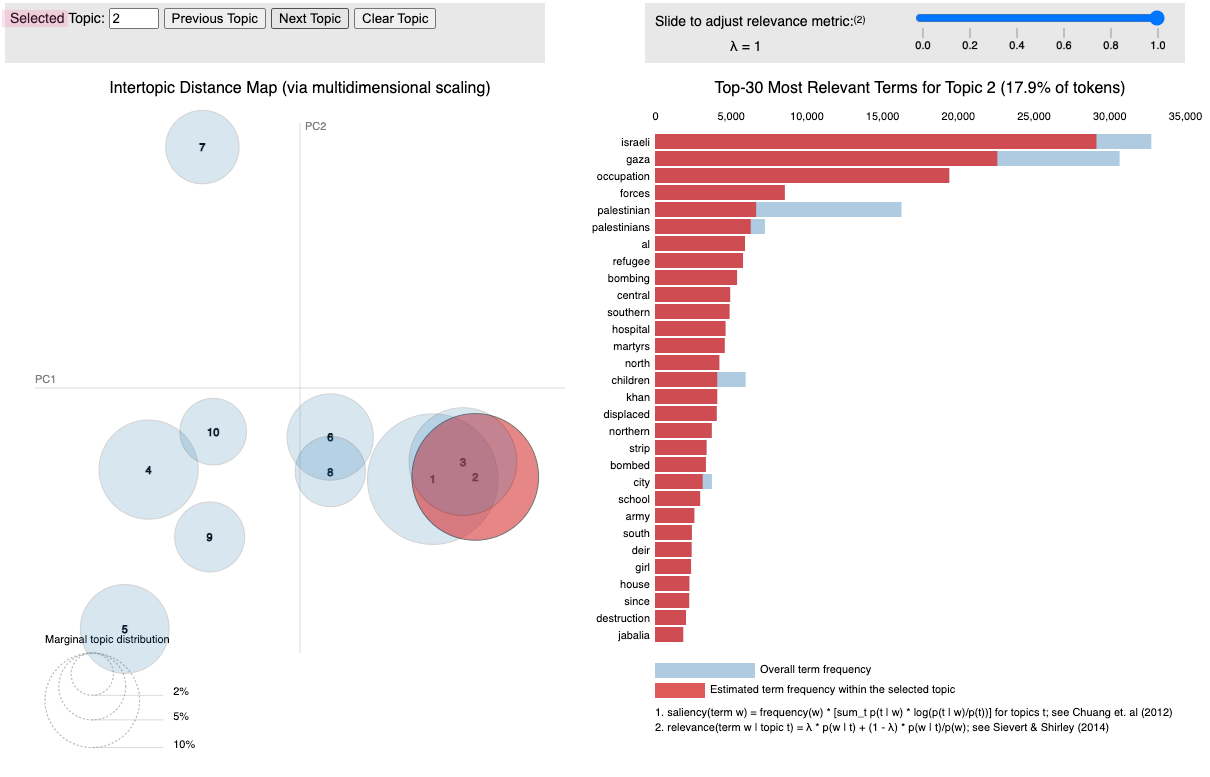}
        \caption{Entities in topic 2}
        \label{fig:topic2_lda}
    \end{subfigure}

    \vspace{0.5cm} % Add some vertical space between rows

    \begin{subfigure}{0.45\linewidth}
        \centering
        \includegraphics[width=\linewidth]{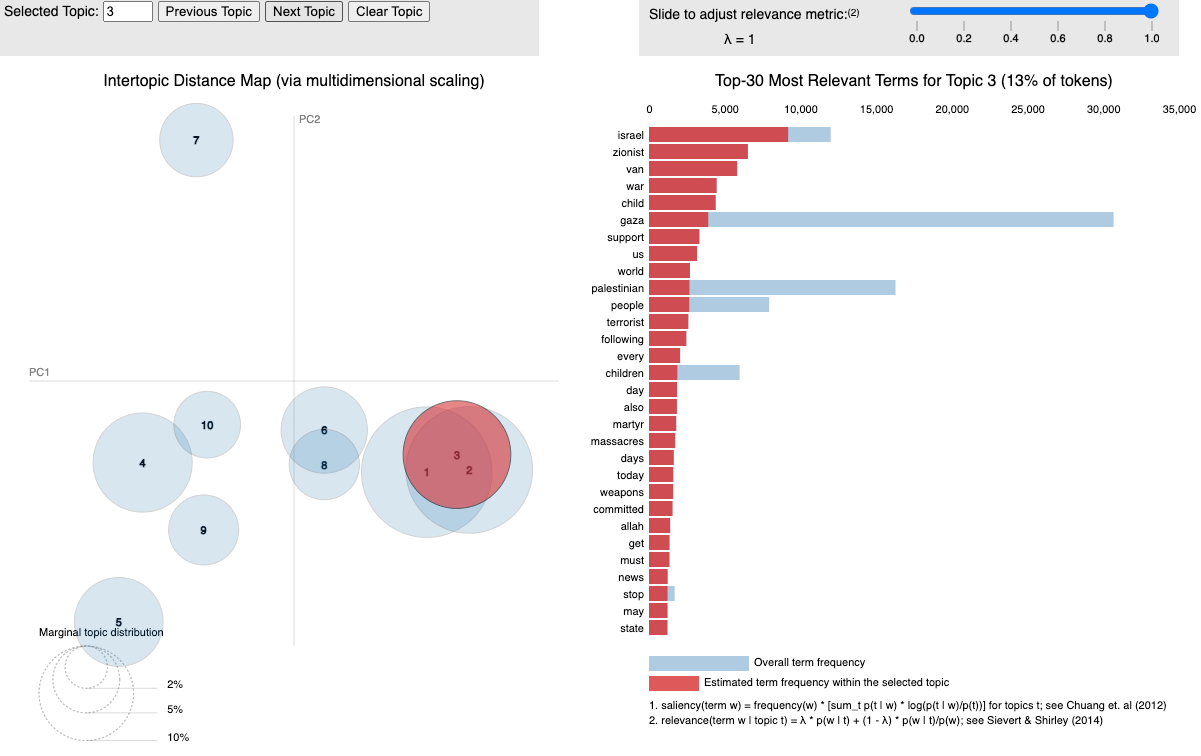}
        \caption{Entities in topic 3}
        \label{fig:topic3}
    \end{subfigure}
    \hfill
    \begin{subfigure}{0.45\linewidth}
        \centering
        \includegraphics[width=\linewidth]{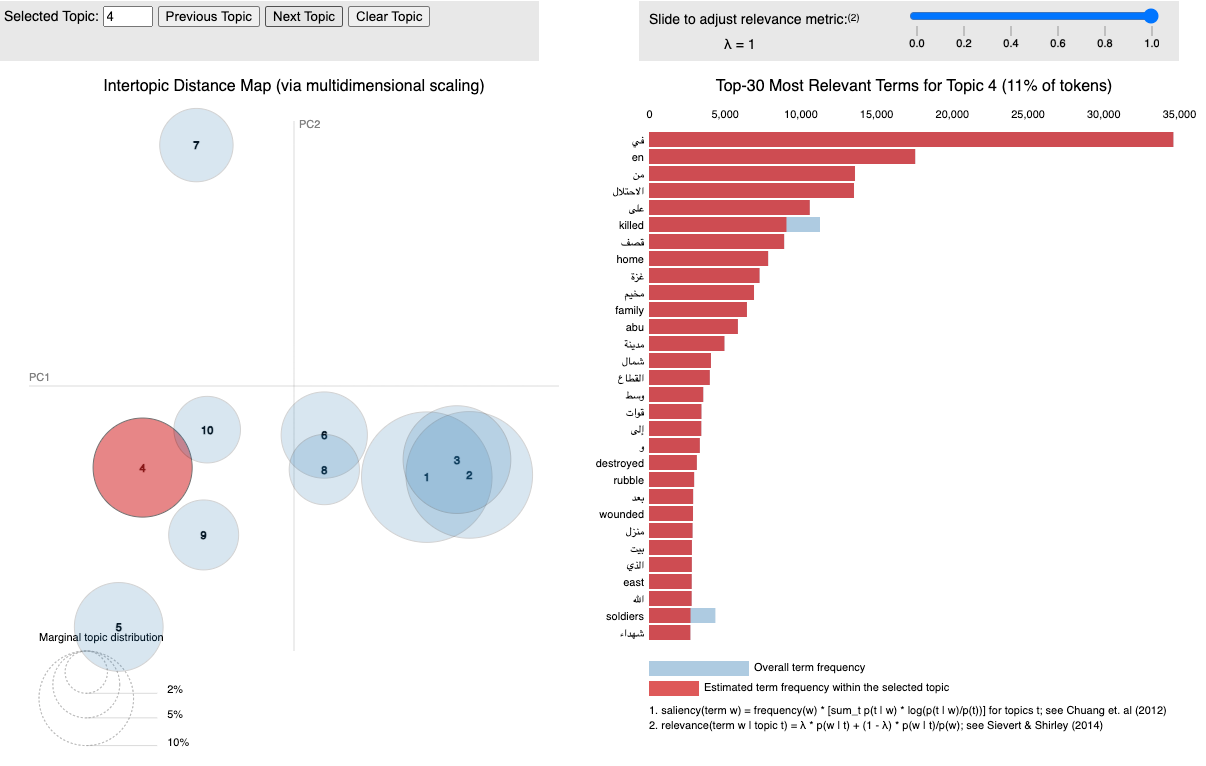}
        \caption{Entities in topic 4}
        \label{fig:topic4_lda}
    \end{subfigure}

    \caption{First 4 topics(1-4)on Telegram from LDA topic analysis}
    \label{fig:lda1}
\end{figure}

\begin{figure}[H]
    \centering
    \begin{subfigure}{0.45\linewidth}
        \centering
        \includegraphics[width=\linewidth]{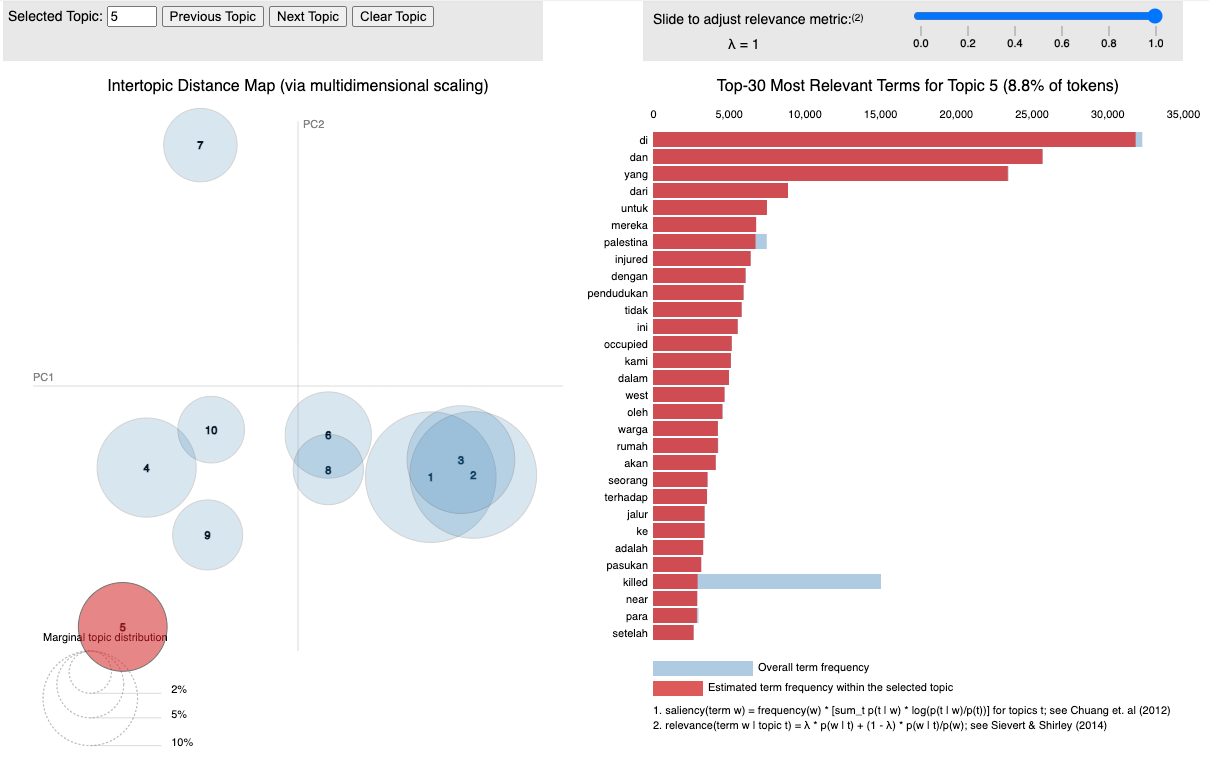}
        \caption{Entities in topic 5}
        \label{fig:topic5_lda}
    \end{subfigure}
    \hfill
    \begin{subfigure}{0.45\linewidth}
        \centering
        \includegraphics[width=\linewidth]{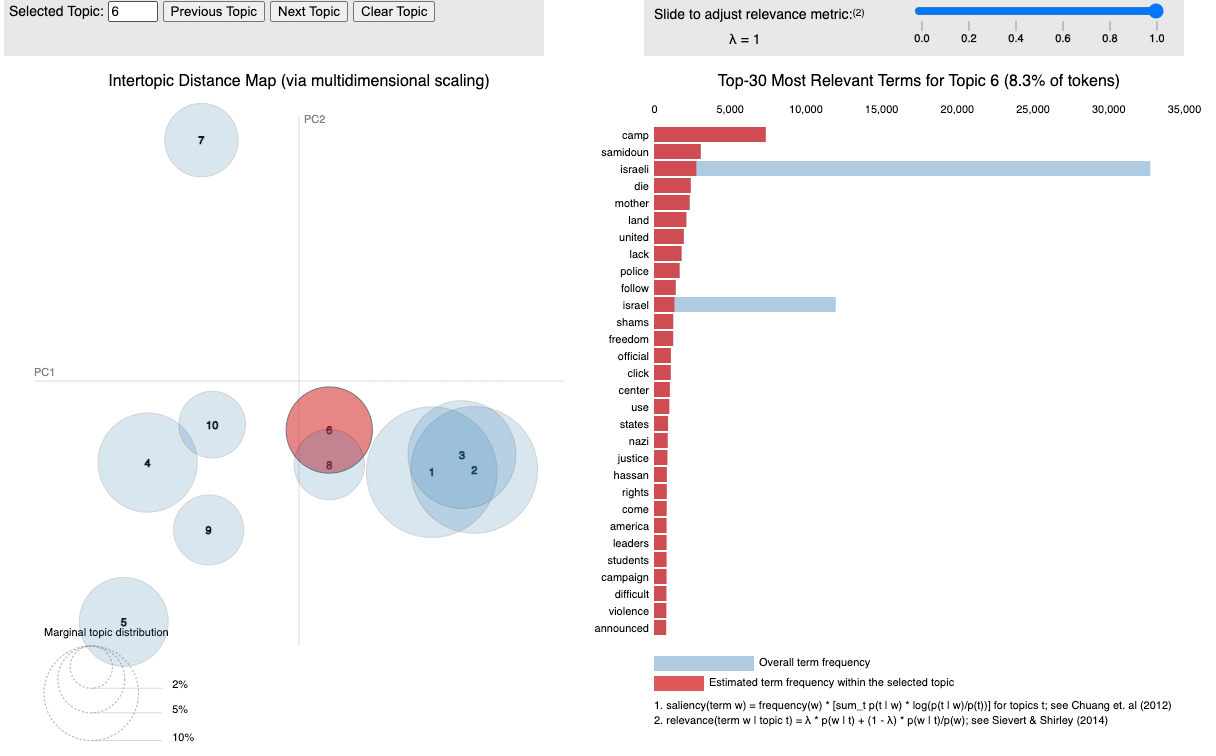}
	    \caption{Entities in topic 6}
        \label{fig:topic6_lda}
    \end{subfigure}

    \vspace{0.5cm} % Add some vertical space between rows

    \begin{subfigure}{0.45\linewidth}
        \centering
        \includegraphics[width=\linewidth]{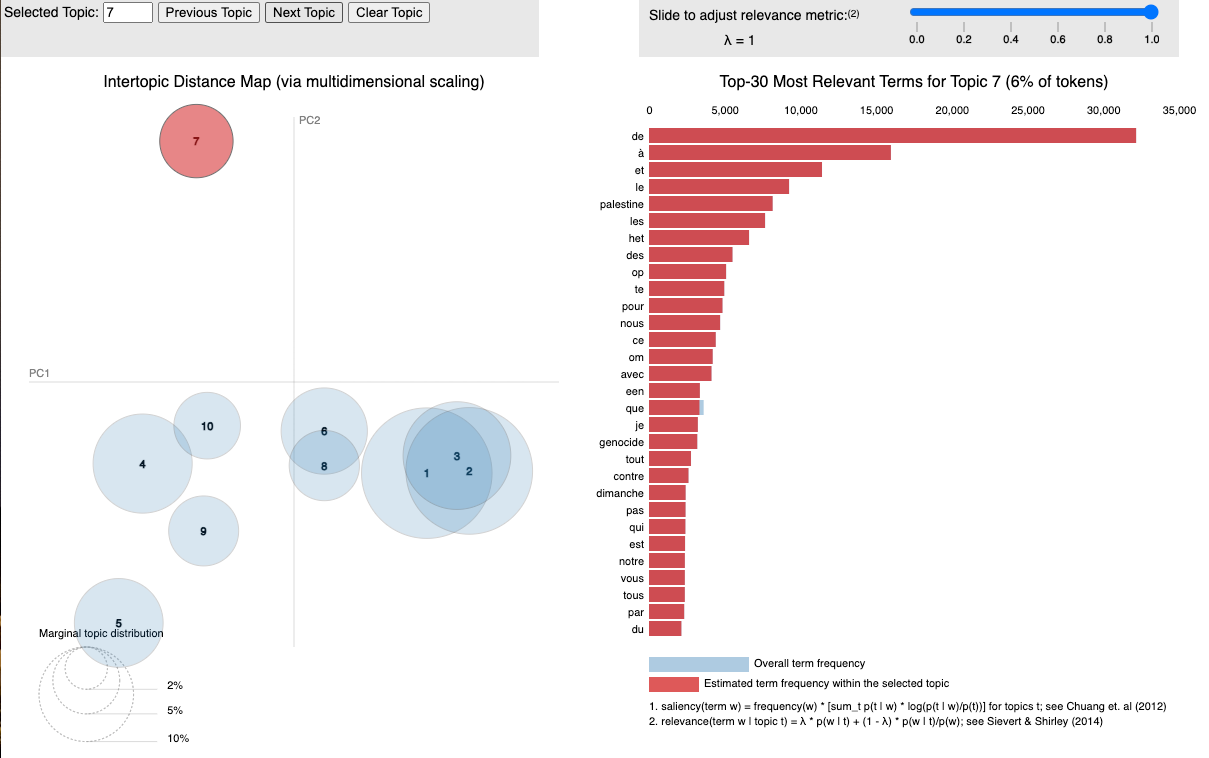}
        \caption{Entities in topic 7}
        \label{fig:topic7_lda}
    \end{subfigure}
    \hfill
    \begin{subfigure}{0.45\linewidth}
        \centering
        \includegraphics[width=\linewidth]{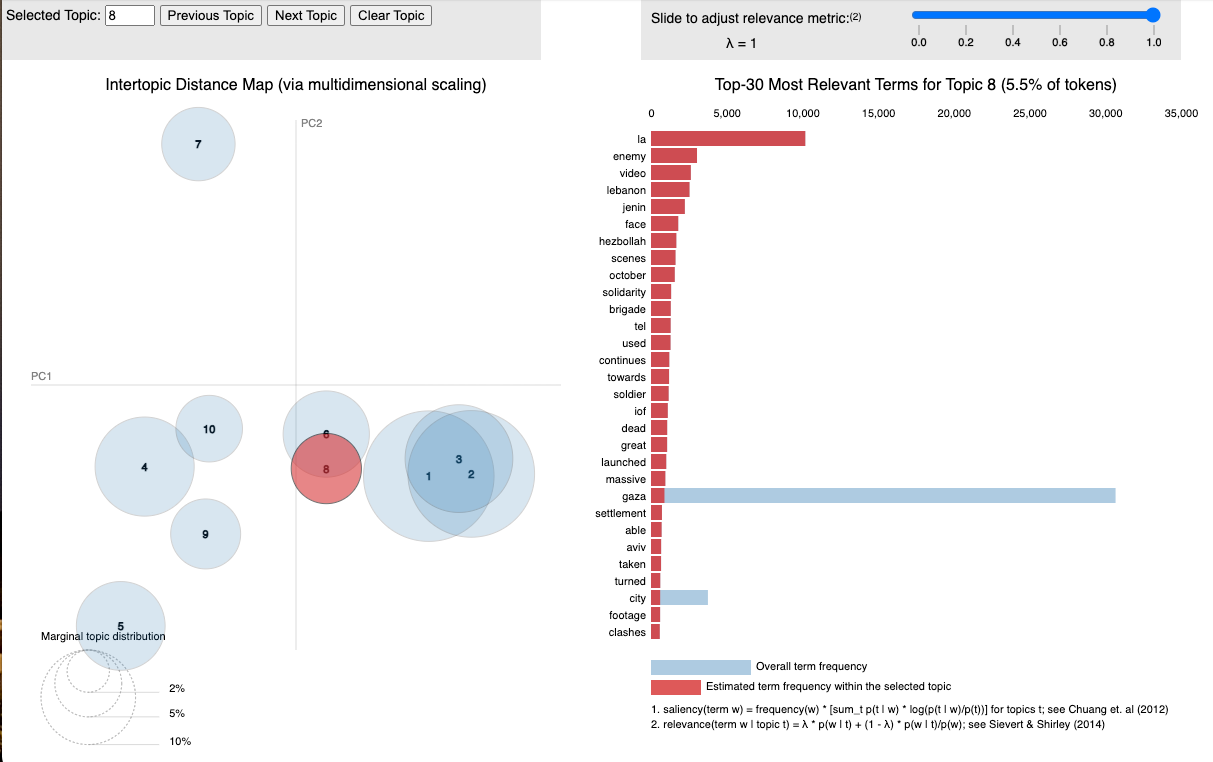}
        \caption{Entities in topic 8}
        \label{fig:topic8_lda}
    \end{subfigure}

    \caption{The next 4 topics(4-8) on Telegram}
    \label{fig:lda2}
\end{figure}

The first step includes the preprocessing text, and then we utilizes the pyLDAvis library for an interactive HTML visualization of the LDA model results, allowing for an in-depth exploration of the identified topics.
The result show the following 8 topics on Telegram illustrated in figure \ref{fig:lda1} and figure \ref{fig:lda2}.

As shown in figure \ref{fig:lda1}, in Topic 1 (19\% of tokens) appears to focus on general discourse about Palestinian issues, with key terms including: "palestinian", "resistance", "people" as primary terms
Media-related terms ("media", "telegram", "link") and action-oriented terms ("targeting", "strikes", "ongoing"). 
The topic's high token percentage (19\%) suggests it represents a major theme in the dataset, while topic 2 (17.9\% of tokens) centers on geographical and humanitarian aspects including a high frequency of location terms ("gaza", "israeli", "northern", "southern"),  Infrastructure-related terms ("hospital", "school", "house")
and humanitarian terms ("refugee", "displaced", "children"). There is a 
strong overlap with Topic 1 in the intertopic distance map, suggesting related discussions. In topic 3 (13\% of tokens) seems to focus on broader international discourse including perspective terms ("world", "support", "us")
Political terms ("zionist", "state"),  Temporal markers ("day", "today", "every").  The intertopic distance map shows Topics 1, 2, and 3 clustering closely together, suggesting these represent interrelated aspects of the main discourse. Together, these three topics account for nearly 50\% of the tokens in the dataset, indicating they represent the core themes in the communications.

 As show in figure \ref{fig:lda2} looking at the intertopic distance maps and term frequency distributions across topics 6-10, we can observe several distinct themes emerging: Topic 7 contains predominantly French language terms ("le", "et", "pour", "nous") alongside "palestine" and "genocide", suggesting French-language discourse. Topic 8 shows terms related to regional actors and locations ("lebanon", "hezbollah", "gaza"). Topic 9 contains military-related terms ("strike", "military", "tentara", "operation"), while Topic 10 focuses on impacts with terms like "air", "targeted", "civilians" and "injuries". The topic distributions and their spatial relationships in the 2D projection suggest these represent distinct but interrelated aspects of discussion. The marginal topic distributions indicate these topics collectively represent about 5-8.3\% of the total tokens in the dataset.

\subsubsection{Bert Topic Model}
%/home/antonakd/TelegramCollector/mamba\_topic/topics\_mamba2.py 
However conventional methods like LDA and NMF, possess some drawbacks due to their manual process of specifying topics in advance and other lacking capabilities. By contrast, modern transformers %\cite{vaswani2017attention} 
have revolutionized the NLP landscape and taken topic modeling to the next level with semantic relationship capturing possible with models such as BERTopic. This algorithm uses embeddings from BERT in a combination with class-based TF-IDF for the generation of interpretable and coherent topics; hence, it is a very powerful tool in practical applications for document analysis.
In general, BERTopic simplifies topic modeling by efficiently identifying topics and their probabilities within documents.
Considering this we proceed with comprehensive textual data analysis using methods of natural language processing combined with machine learning, using BERTopic.

We use the Bert topic library applied in topic modeling that enables generating topics from text by making use of transformers and traditional machine learning methods such as CountVectorizer. The CountVectorizer, a scikit-learn tool is used to transform our dataset into a matrix of token count. The SentenceTransformer (MiniLM), a transformer-based model is loaded for the embedding of text. We process the corpus by initially cleaning the text (removing URLs, special characters, and lowercasing), and store the cleaned messages in a pandas DataFrame. The train\_model() method uses BERTopic, a topic modeling library, to discover topics from the messages using a count-based vectorizer that considers n-grams. The analyze\_topics() method analyzes the discovered topics, extracting key words for each topic and calculating topic distribution across different channels and over time.
 
\begin{figure}[H]
	\centering
	\includegraphics[width=0.9\linewidth]{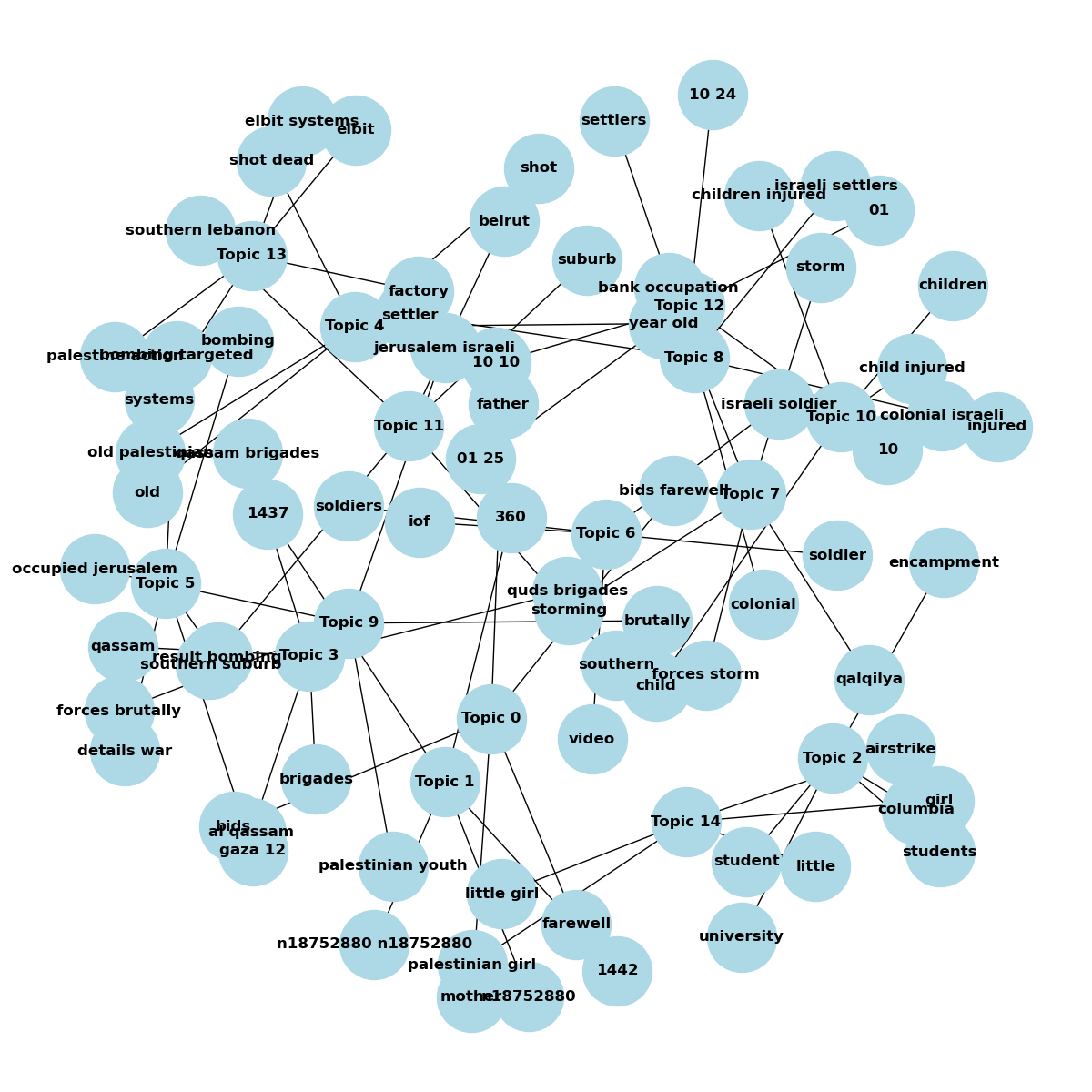}
	\caption{A network of connections between the top ten most important topics on Telegram}
	\label{top_ten_topics_bert}
\end{figure}

\begin{figure}[H]
    \centering
    \includegraphics[width=0.9\linewidth]{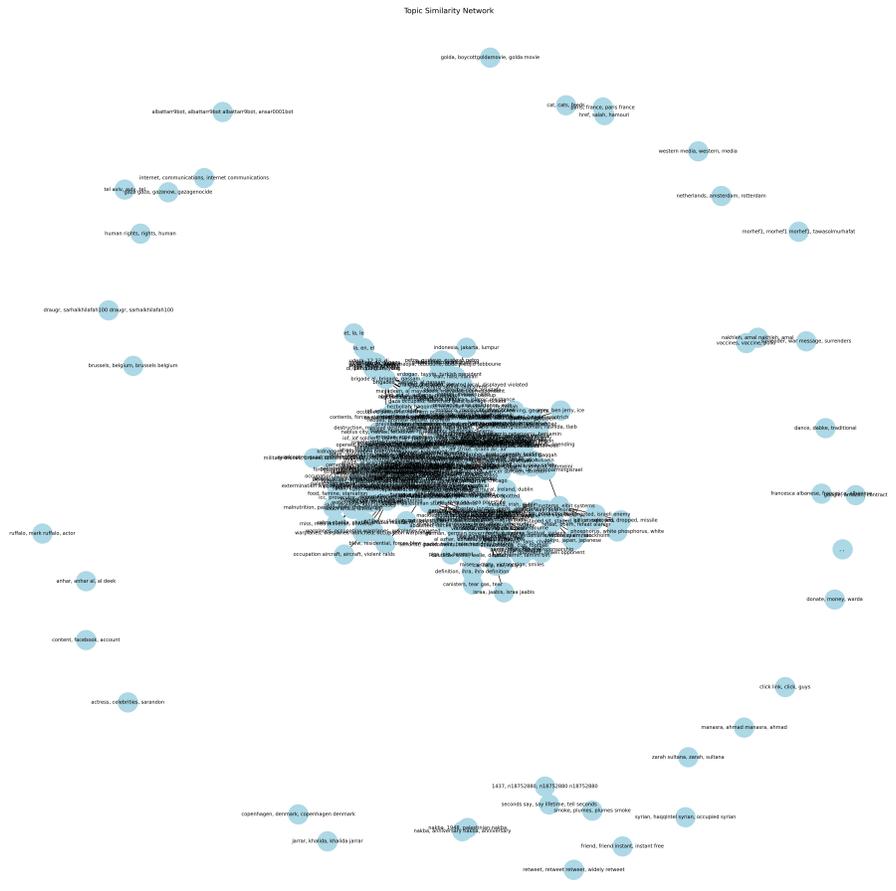}
    \caption{Network of 309 topics on Telegram}
    \label{topicnet}
\end{figure}

\begin{figure}[H]
    \centering
    \includegraphics[width=0.9\linewidth]{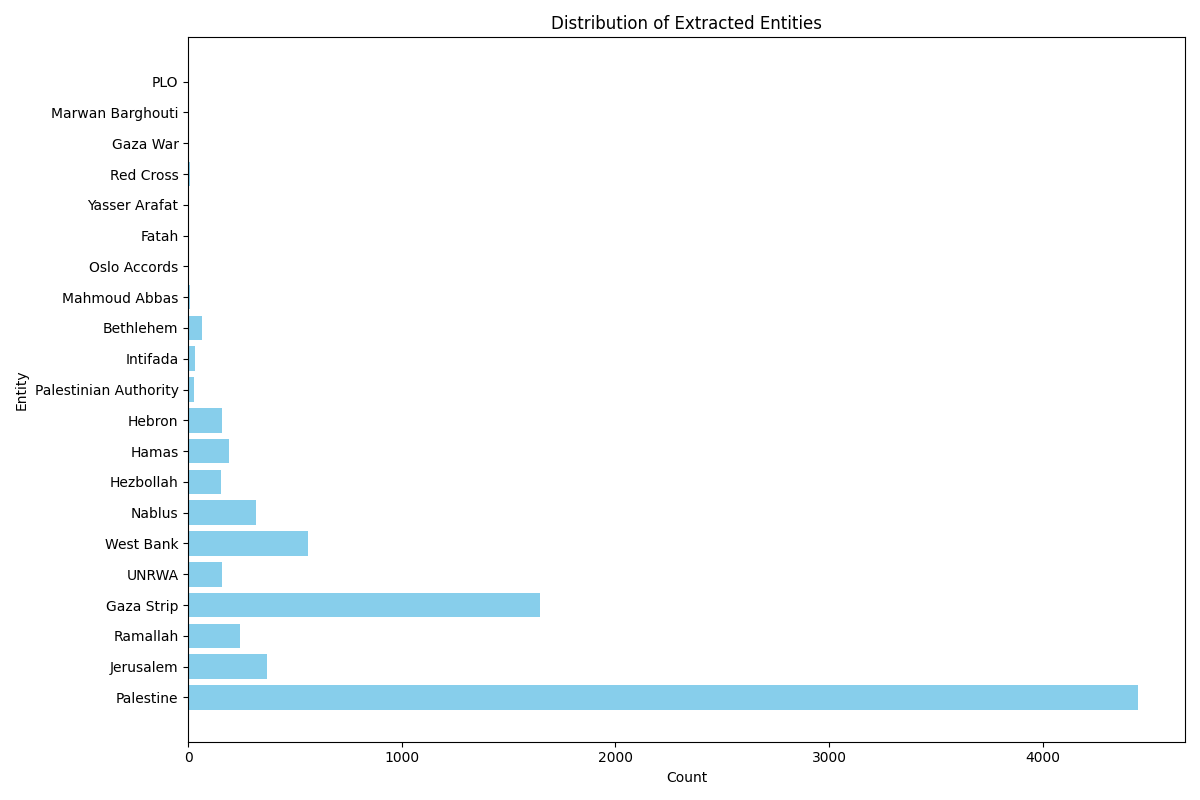}
    \caption{Distribution of entities by applying Bert topic analysis, on the number of messages  }
    \label{fig:entities_bert}
\end{figure}

Specifically, we analyze 51403 out of 70321 messages because we remove duplicates and non-text messages, from 17 channelsfrom telegram, and proceed to the training of the topic model, transforming documents to embeddings, using PyTorch cuda %\cite{pytorch_cuda}. 
Next we load the pretrained SentenceTransformer: sentence-transformers/all-MiniLM-L6-v2 with 1607 batches:. 
Finally, we fit to the dimensionality reduction algorithm and proceed to the clustering of the the reduced embeddings. 
The resulted topics discovered were 309 with results shown in fig \ref{topicnet}.
A graph with the top ten most important nodes are shown in figure \ref{top_ten_topics_bert}. This visualization shows the relationships and associations of various themes and entities to the Israeli-Palestinian conflict. The recurring themes are military operations, civilian impact, political actors, and geographic places. The highly interconnected nature of these subjects speaks to the complexity underlying this long-standing conflict, argued across digital platforms. We notice the Israeli settlers that relate to such themes as colonial Israeli injured and children injured.
Palestinian youth relate to the theme of farewel and a Palestinian girl.Quds brigades and Gassam brigades are related to themes about brigades, southern forces, storming.
Elbit systems - defense firm relates to a theme called "elbit systems shot dead". Geographical locations: Southern Lebanon, Beirut, Gaza also relates to several themes concerning the conflicts.
This graph in its entirety depicts the interwoven relationships between military actors, civilian impacts, political elements, and geographic areas involved in this long-standing conflict.
Finally, we show the distribution of top entities per topics in figure \ref{fig:entities_bert}.

\begin{comment}
\begin{figure}[H]
    \centering
    \includegraphics[width=0.9\linewidth]{mamba/topic\_heatmap.png}
    \caption{Topic analysis on Telegram form 309 topics heatmap}
    \label{fig:topicheat_mamba}
\end{figure}
\end{comment}

\begin{comment}
    \begin{figure}[H]
    \centering
    \includegraphics[width=0.9\linewidth]{mamba/topic\_importance.png}
    \caption{Topic importance for 309 topics}
    \label{fig:importantce_mamba}
\end{figure}
\end{comment}

\iffalse 
\begin{table}[H]
    \centering
    \begin{tabular}{|c|l|}
        \hline
        \textbf{Entity Type} & \textbf{Entities} \\
        \hline
        LOCATION & Palestine, West Bank, Gaza Strip, \\ 
                  & Jerusalem, Hebron, Nablus, \\ 
                  & Ramallah, Bethlehem \\
        \hline
        ORGANIZATION & PLO, Hamas, Fatah, \\ 
                     & Palestinian Authority, Hezbollah, \\ 
                     & UNRWA, Red Cross \\
        \hline
        PERSON & Yasser Arafat, Mahmoud Abbas, \\ 
               & Hanan Ashrawi, Salah Khalaf, \\ 
               & Marwan Barghouti, Nadia Hijab \\
        \hline
        EVENT & Israeli-Palestinian Conflict, \\ 
              & Intifada, Oslo Accords, \\ 
              & Gaza War \\
        \hline
    \end{tabular}
    \caption{Entity Types and Their Associated Keywords}
    \label{tab:entities}
\end{table}
\fi
%The post illustrates the entities defined in the initial step and cluster the data to perform a fine-grained analysis of thematic patterns. Further, we enhance our analysis by applying K-Means clustering on the topics in order to classify these topics and check for coherence using silhouette scores  Finally, we save the results, including visual representations of entity distributions and topic information, for further investigation.  

\subsubsection{Twitter}

\begin{figure}[H]
    \centering
    \includegraphics[width=0.9\linewidth]{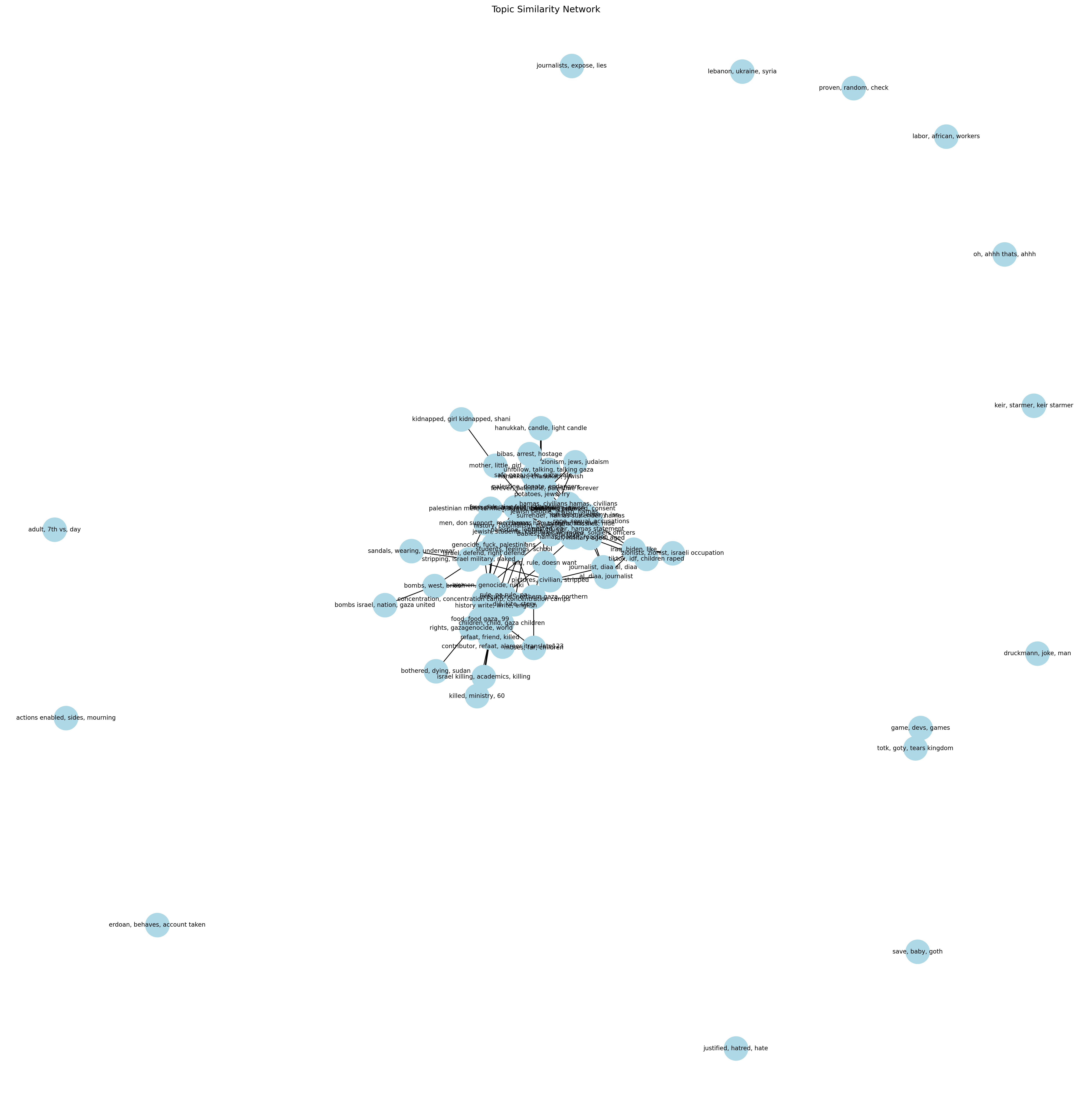}
    \caption{Network of topics for Twitter dataset, contaning 309 topics}
    \label{fig:topictwitter}
\end{figure}

\begin{figure}[H]
    \centering
    \includegraphics[width=0.9\linewidth]{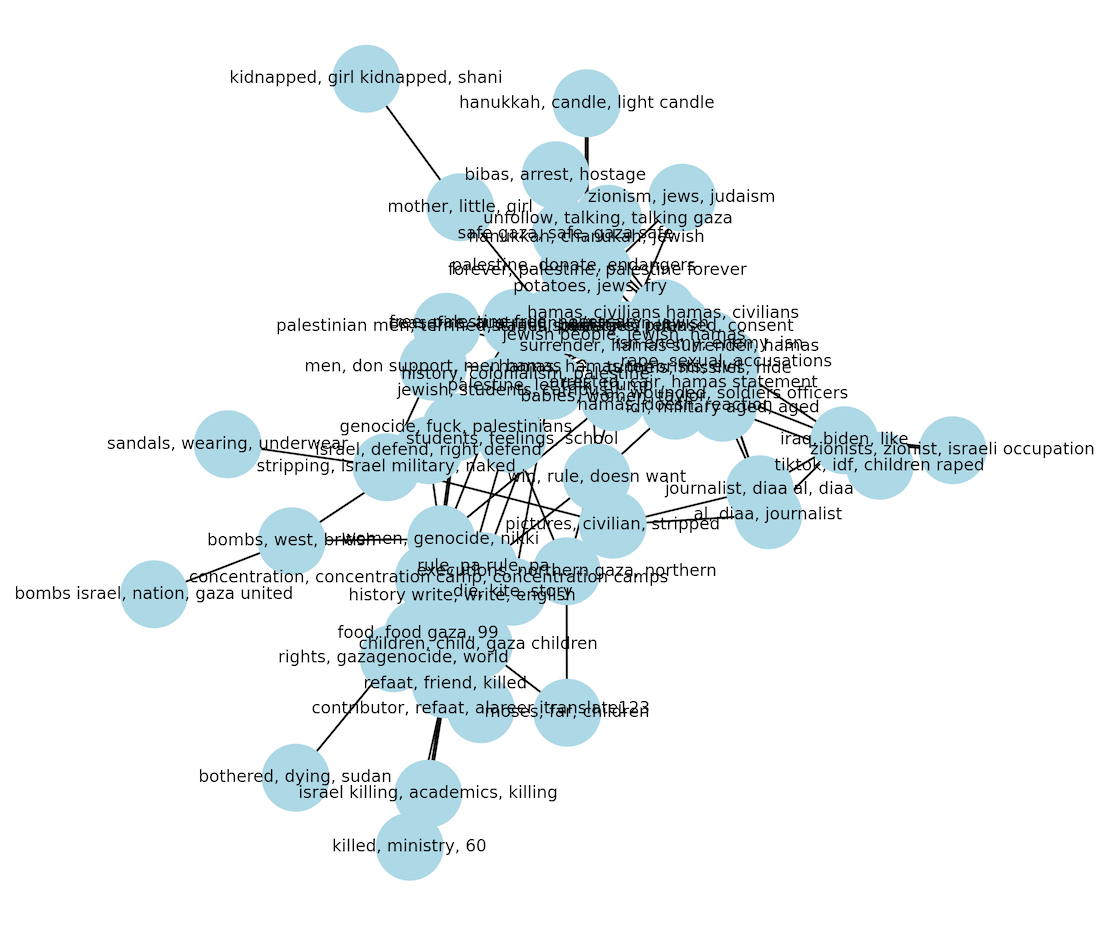}
    \caption{The network of topics for Twitter}
    \label{fig:topictwitter_zoom}
\end{figure}

\begin{figure}[H]
    \centering
    \includegraphics[width=0.9\linewidth]{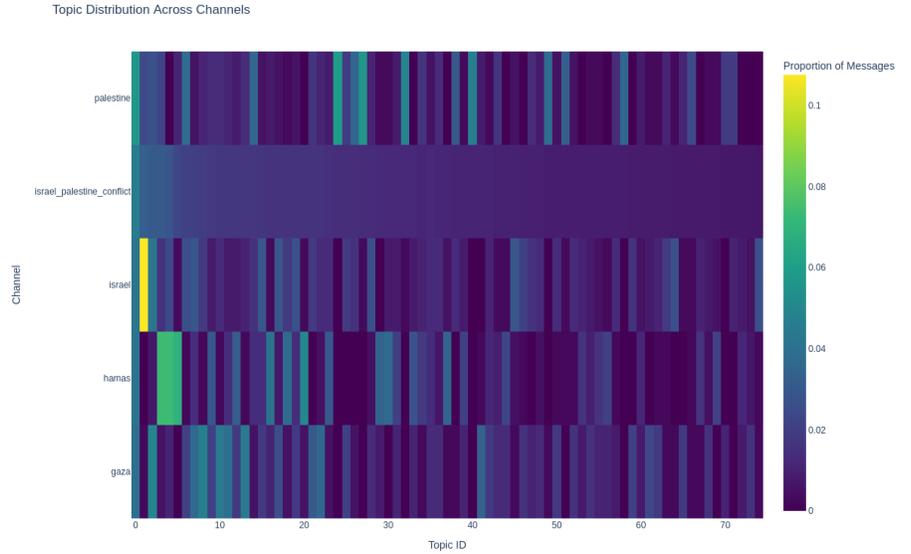}
    \caption{The heatmap network of Twitter topics}
    \label{fig:topictwitter_heatmap}
\end{figure}

We apply the same topic analysis on the Twitter dataset using the BERTopic model. Initially, we preprocess the messages, train a topic model, and generate the most relevant topic. Additionally, weo analyze the topic distribution across different channels and timestamps. The results are shown in the following graphs \ref{fig:topictwitter}, \ref{fig:topictwitter_zoom} and \ref{fig:topictwitter_heatmap}.
% All visualizations have been saved to: /home/antonakd/twitter/rizqikapratamaa/visualizations (base) antonakd@blackmamba:~/twitter/rizqikapratamaa python3 heatmap\_topic.py
In figure \ref{fig:topictwitter} we show a visualization of the topic similarity network, showing relations between the various topics of discussion on the Israeli-Palestinian conflict. 
It looks dominated by terms about the military features of the conflict: "brigades", "soldiers", and "storming", among geographic places such as "Jerusalem" and "Gaza".  From the above, it follows that operational details in general and in regard to military operations are stressed.
Another cluster is very different, with more humanitarian and civilian issues, including phrases like "children injured", "southern forces", and "occupied Jerusalem". This would show discussions of the human impact and civilian suffering caused by the conflict.
More thematic sets are also related to wider geopolitical themes, such as "farewell", "palestinian girl", and further some international actors and media. It shows that the conflict is debated within a framing of the wide regional and global context. The network visual representation really depicts an overview of all military, humanitarian, and political features that are involved and represented with respect to the discussion on the Israeli-Palestinian conflict in the form of such a massive web about a long and highly debated discussion.

\section{Sentiment Analysis}
\subsection{Twitter}
\begin{figure}[H]
    \centering
    \begin{subfigure}[b]{0.45\linewidth}
        \centering
        \includegraphics[width=\linewidth]{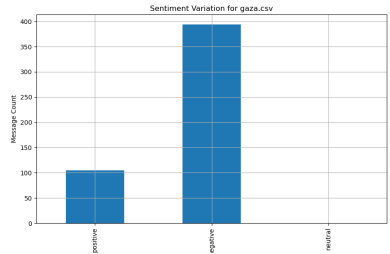}
        \caption{Sentiments for }
        \label{fig:emotion_plot1_twitter}
    \end{subfigure}
    \hfill
    \begin{subfigure}[b]{0.45\linewidth}
        \centering
        \includegraphics[width=\linewidth]{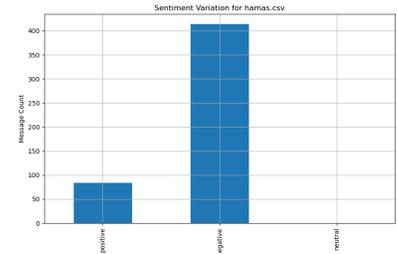}
        \caption{Sentiments for GazaEnglishUpdates}
        \label{fig:emotion_plot2_twitter}
    \end{subfigure}
    
    \vspace{1em}  % Add space between rows
    \begin{subfigure}[b]{0.45\linewidth}
        \centering
        \includegraphics[width=\linewidth]{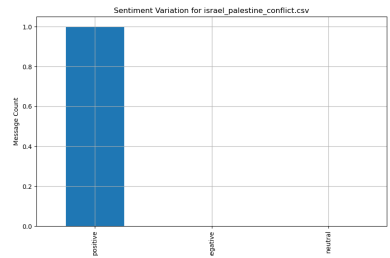}
        \caption{Sentiments for GazaNow}
        \label{fig:emotion_plot3_twitter}
    \end{subfigure}
    \hfill
    \begin{subfigure}[b]{0.45\linewidth}
        \centering
        \includegraphics[width=\linewidth]{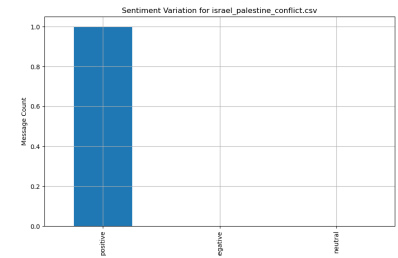}
        \caption{Sentiments for Pal\_Online9}
        \label{fig:emotion_plot4_twitter}
    \end{subfigure}
    
    \caption{Granular Sentiment across Various Channels}
    \label{fig:granular_emotion_plot_combined_twitter}
\end{figure}

We perform sentiment analysis on our Twitter corpus using the Hugging Face transformers library. 
We process each message to determine whether its sentiment is positive, negative, or neutral. Next we generates the bar plots visualizing the sentiment distribution shown in \ref{fig:emotion_plot1_twitter}, \ref{fig:emotion_plot2_twitter},\ref{fig:emotion_plot3_twitter} and \ref{fig:emotion_plot4_twitter}. 

%/home/antonakd/twitter/rizqikapratamaa mamba\_sentiment8.py
 
\subsection{Telegram}

The final step is to perform  granular sentiment analysis in order to reveal specific emotions like hope, support, anger, and despair. Towards this goal we incorporate a model that can classify texts into more detailed emotion categories. Hugging Face provides models like j-hartmann/emotion-english-distilroberta-base, which classifies text into multiple emotion categories rather than just positive, negative, and neutral. 

 \begin{figure}[H]
    \centering
    \begin{subfigure}[b]{0.45\linewidth}
        \centering
        \includegraphics[width=\linewidth]{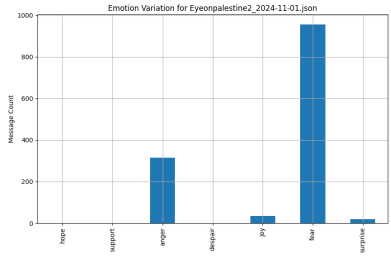}
        \caption{Sentiments for Eyeonpalestine2}
        \label{fig:emotion_plot1}
    \end{subfigure}
    \hfill
    \begin{subfigure}[b]{0.45\linewidth}
        \centering
        \includegraphics[width=\linewidth]{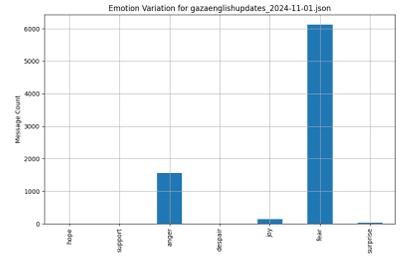}
        \caption{Sentiments for GazaEnglishUpdates}
        \label{fig:emotion_plot2}
    \end{subfigure}
    
    \vspace{1em}  % Add space between rows
    \begin{subfigure}[b]{0.45\linewidth}
        \centering
        \includegraphics[width=\linewidth]{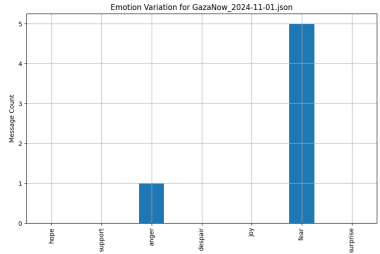}
        \caption{Sentiments for GazaNow}
        \label{fig:emotion_plot3}
    \end{subfigure}
    \hfill
    \begin{subfigure}[b]{0.45\linewidth}
        \centering
        \includegraphics[width=\linewidth]{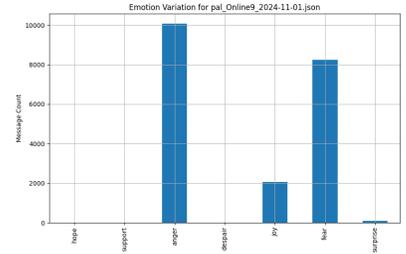}
        \caption{Sentiments for Pal\_Online9}
        \label{fig:emotion_plot4}
    \end{subfigure}
    
    \caption{Granular Sentiment across Various Channels}
    \label{fig:granular_emotion_plot_combined}
\end{figure}

We apply a pre-trained emotion analysis model from Hugging Face's transformers library. The dominant emotion for each non empty, English, message is identified to finally generate specific emotion analysis as shown in bar plots \ref{fig:emotion_plot1}, \ref{fig:emotion_plot2} \ref{fig:emotion_plot3} \ref{fig:emotion_plot4}. In these plots we show the distribution of emotions across the messages and notice that the dominant emotions as expected are anger and fear.  

\subsection{Reddit}

\begin{figure}[H]
	\centering
	\includegraphics[width=0.9\linewidth]{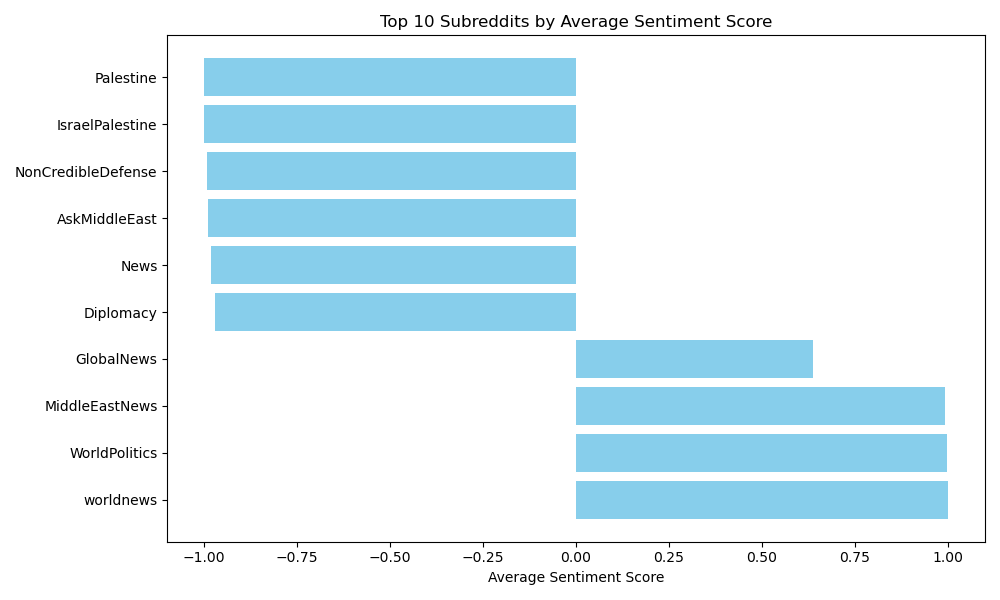}
    	\caption{Sentiment for Reddit}
	\label{reddit_sent}
\end{figure}

In order to apply sentiment analysis on our reddit corpus we load a pre-trained sentiment analysis model from HuggingFace and use it to analyze the sentiment of text data from various subreddits. It processes a dataframe, calculating sentiment scores for each text and averaging the sentiment per subreddit. Then, it creates a horizontal bar plot for the top 10 subreddits by average sentiment score and saves the plot as an image file. The results are shown in figure \ref{reddit_sent}.
\section{Conclusion}
The discussion of the war between Israel and Hamas in the current study provides the first comprehensive insight, while additional comparison was also drawn from Twitter and Reddit. An analysis of 105,000 messages on Telegram, 2,001 tweets, and 2 million opinions sourced from Reddit, several patterns materialized during our analysis with great importance to this online discursive exchange.
Frequency analysis showed the volume of the messages increased dramatically after events on October 7, 2023; this is mainly reflected in several channels. Following this, entity extraction identified the key themes and players.
Advanced NLP techniques applied herein include topic modeling, which, through means such as LDA and BERTopic analysis, demonstrated eight major topic clusters on humanitarian concerns to military operations. Our topic modeling showed that 19\% of the discussions talked about general Palestinian issues; 17.9\% of the discussions were about geographical and humanitarian aspects, while the remaining ones were composed of international discourses, French language communications, regional actors, and military operations.
SA showed highly polarized emotional responses across platforms. Granular emotion analysis, in particular, presented that the dominant emotions expressed in messages, especially in channels focused on direct conflict reporting, are those of anger and fear. This emotive polarization strongly seems to correlate with specific topics and events. Hence, information sharing may be in complex interaction with emotive responses.
Key implications:

Platform Dynamics: Indeed, Telegram is used as a main communication platform in a conflict situation; this however is subject to considerable limitations through content moderation efforts.
Information Flow: The study shows how different narratives evolve and spread across channels, with some topics coming to the fore based on real-world events.
Emotional Impact: The dominance of negative sentiments, especially that of anger and fear, calls for an urgent requirement of understanding how digital platforms shape conflict perception and emotional response.
 Methodological Contributions: Combined used of LDA and BERTopic analysis thus offers a robust framework in analyzing complex and multilingual discourse during ongoing conflicts.

Future studies could thus be informed by longer-term trends, the impact of content moderation on information flow, and more sophisticated methods for detecting and analyzing propaganda efforts. Moreover, comparative studies across different conflicts might establish whether the patterns observed here are peculiar to this conflict or representative of broader trends in social media communication during times of crisis.

\bibliography{main}

\section{Appendix}

\subsection{Additional volume plots}
\label{Add_volume}

\begin{figure}[H]
    \centering
    \begin{subfigure}{0.45\linewidth}
        \centering
        \includegraphics[width=\linewidth]{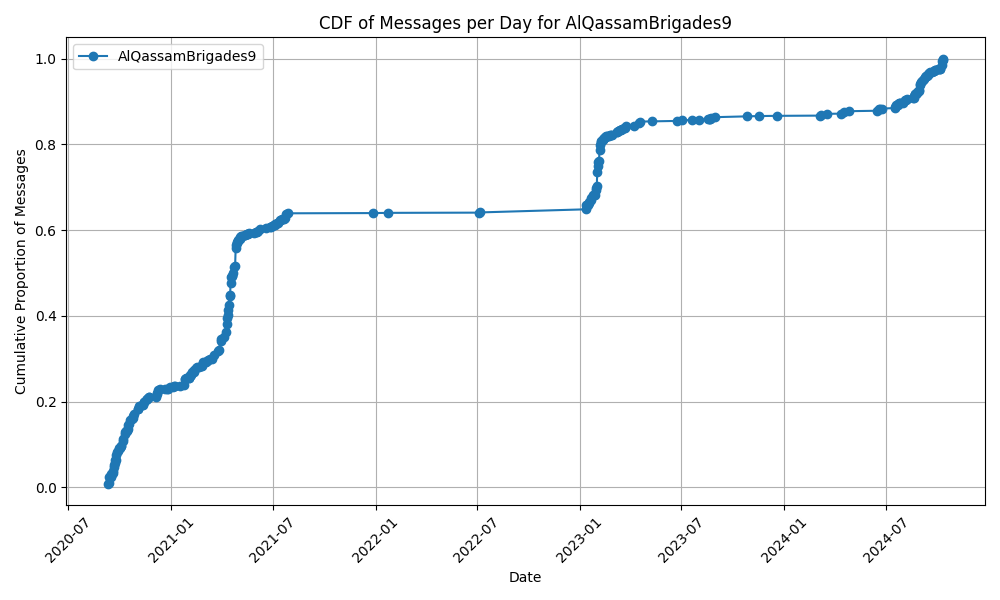}
        \caption{Cumulative distribution function of messages in 'Al Qassam Brigades' channel}
        \label{fig:AlQassamBrigades9_cdf_plot}
    \end{subfigure}
    \hfill
    \begin{subfigure}{0.45\linewidth}
        \centering
        \includegraphics[width=\linewidth]{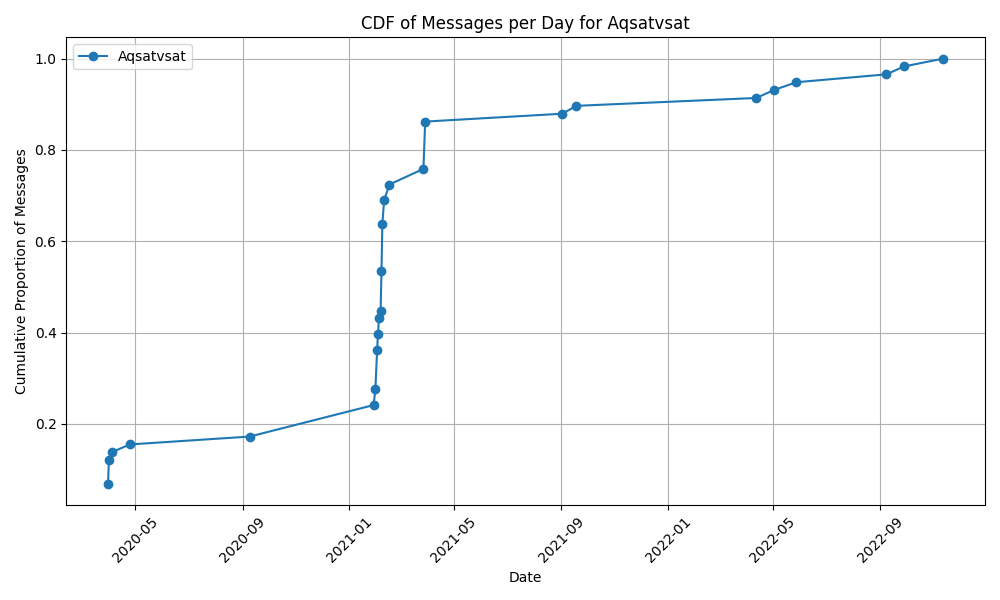}
        \caption{Cumulative distribution function of messages in 'Aq satvsat' channel}
        \label{fig:Aqsatvsat_cdf_plot}
    \end{subfigure}

    \vspace{0.5cm} % Add some vertical space between rows

    \begin{subfigure}{0.45\linewidth}
        \centering
        \includegraphics[width=\linewidth]{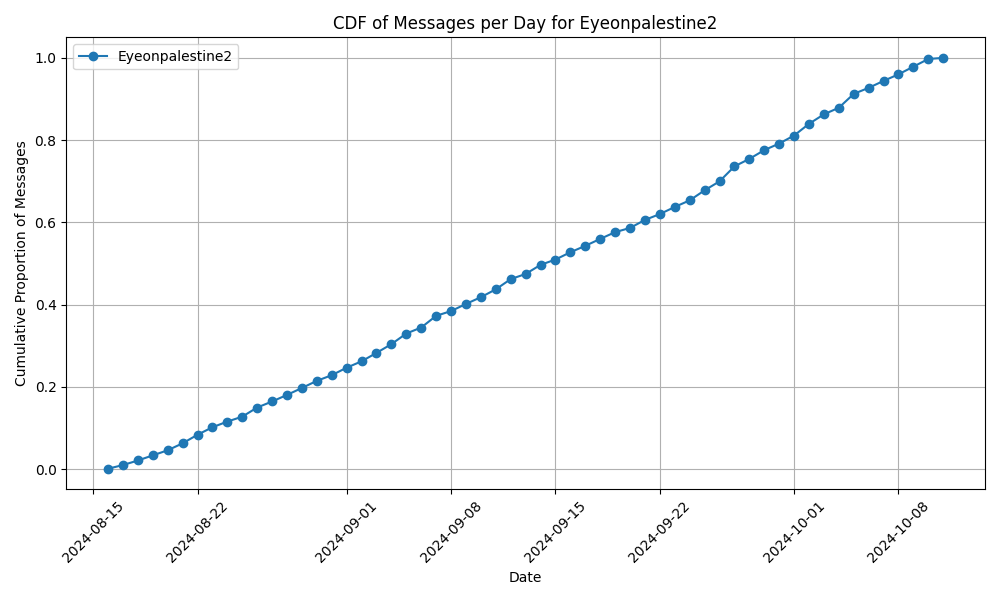}
        \caption{Caption for Topic 7}
        \label{fig:Eyeonpalestine2_cdf_plot}
    \end{subfigure}
    \hfill

    \caption{Overall Caption for All Topics}
    \label{fig:overall_all_topics}
\end{figure}

\end{document}